\documentclass[%
 reprint,
superscriptaddress,
nofootinbib,
showkeys,
 amsmath,amssymb,
 aps,
]{revtex4-2}

\usepackage{graphicx}
\usepackage{dcolumn}
\usepackage{bm}
\usepackage[labelformat=simple]{subcaption}

\usepackage[dvipsnames]{xcolor}
\usepackage{soul}
\usepackage{float}
\usepackage{multirow}

\usepackage{enumitem,amssymb}
\newlist{todolist}{itemize}{2}
\setlist[todolist]{label=$\square$}
\usepackage{pifont}
\usepackage[colorlinks=true,linkcolor=MidnightBlue, citecolor=DarkOrchid,urlcolor=PineGreen]{hyperref}

\begin{document}

\preprint{APS/123-QED}

\title{Motifs in earthquake networks:\\Romania, Italy, United States of America, and Japan}

\author{Gabriel Tiberiu Pan\u{a}}
    \email{gabriel.pana@s.unibuc.ro}
    \affiliation{Faculty of Physics, University of Bucharest, Atomi\c{s}tilor 405, M\u{a}gurele, Rom\^{a}nia}

\author{Alexandru Nicolin-\.{Z}aczek}%
    \email{alexandru.nicolin@spacescience.ro}
    \affiliation{Institute of Space Science, Atomi\c{s}tilor 409, M\u{a}gurele, Rom\^{a}nia}
    \thanks{formerly Alexandru Nicolin}

\date{\today}

\begin{abstract}
We present a detailed description of seismic activity in Romania, Italy, and Japan, as well as the California seismic zone in the United States of America, based on the statistical analysis of the underlying earthquake networks used to model the aforementioned zones. Our results on network connectivity and simple network motifs allow for a complex description of seismic zones, while at the same time reinforcing the current understanding of seismicity as a critical phenomenon. The reported distributions on node connectivity, three-, and four-event motifs are consistent with power-law, {\it i.e.}, scale-free, distributions over large intervals and are robust across earthquake networks obtained from different discretizations of the seismic zones of interest. In our analysis of the distributions of node connectivity and simple motifs, we distinguish between the global distribution and the power-law part of it with the help of maximum likelihood estimation (MLE) method and complementary cumulative distribution functions (CCDF). The main message is that the distributions reported for the aforementioned seismic zones have large power-law components, extending over some orders of magnitude, independent of discretization. All the results were obtained using publicly-available databases and open-source software, as well as a new toolbox available on GitHub, specifically designed to automatically analyze earthquake databases. 
\end{abstract}


                              
\maketitle


\section{\label{sec:introduction}Introduction}

The apparent ubiquity of distributions that resemble power laws has effectively turned almost every field of research into a contributor to the science of complex critical systems. While the list of reported power-law-like distributions seems endless, with observations ranging from linguistics \cite{Piantadosi2014}, musicology \cite{Serra-Peralta2022}, and biology \cite{Brown2002} to meteorology \cite{hsu1994}, environmental studies \cite{McKenzie2012}, economics \cite{Gabaix2009}, and computer science \cite{Faloutsos1999}, the list of structural models which explain these distributions is considerably smaller. This asymmetry reflects, on one hand, the large amount of empirical data, while on the other hand, the difficulty of constructing models that exhibit power-law scaling which is caused by the subtle interplay between the components of a given system. On the side of the available empirical data, we recall that the main defining {\it v}'s of Big Data are volume, variety, velocity, and volatility, effectively turning data into a new commodity \cite{Aaltonen2021, Sadowski2019}. This should be contrasted with the substantially fewer available models, for the creation of which there are no clear recipes, because of the numerous challenges which stem from the treatment of the underlying physical processes across different scales of time and space. 

One of the most striking properties of complex systems is the possibility to self-organize into a critical state, a process which has been initially proposed by Bak {\it et al.} in 1987 through a now classical sandpile model \cite{bak1987}. The emergence of scale-free distributions of various observables was the main fingerprint of self-organized criticality (SOC) that numerous subsequent models have tried to reproduce. Among them, we mention here the earthquake model of Olami–Feder–Christensen (OFC) \cite{olami1992}, which exhibited SOC and was able to reproduce empirical laws such as the Gutenberg–Richter law and Omori law. It should be noted that the OFC model draws from numerous older mechanical earthquake models \cite{burridge1967, otsuka1972, Carlson1989}, but none of the previous ones was as versatile or as effective in reproducing empirical laws. The underlying lattice of the initial SOC models was regular, but now many models use networks as well, and it is commonly accepted that the topology of the underlying network can impact the observed dynamics, see Refs. \cite{caruso2006, peixoto2006} for the OFC model over different types of networks.    

\begin{figure*}[hbt!]
    \centering
    \begin{subfigure}[b]{0.49\textwidth}
         \centering
         \includegraphics[width=\textwidth]{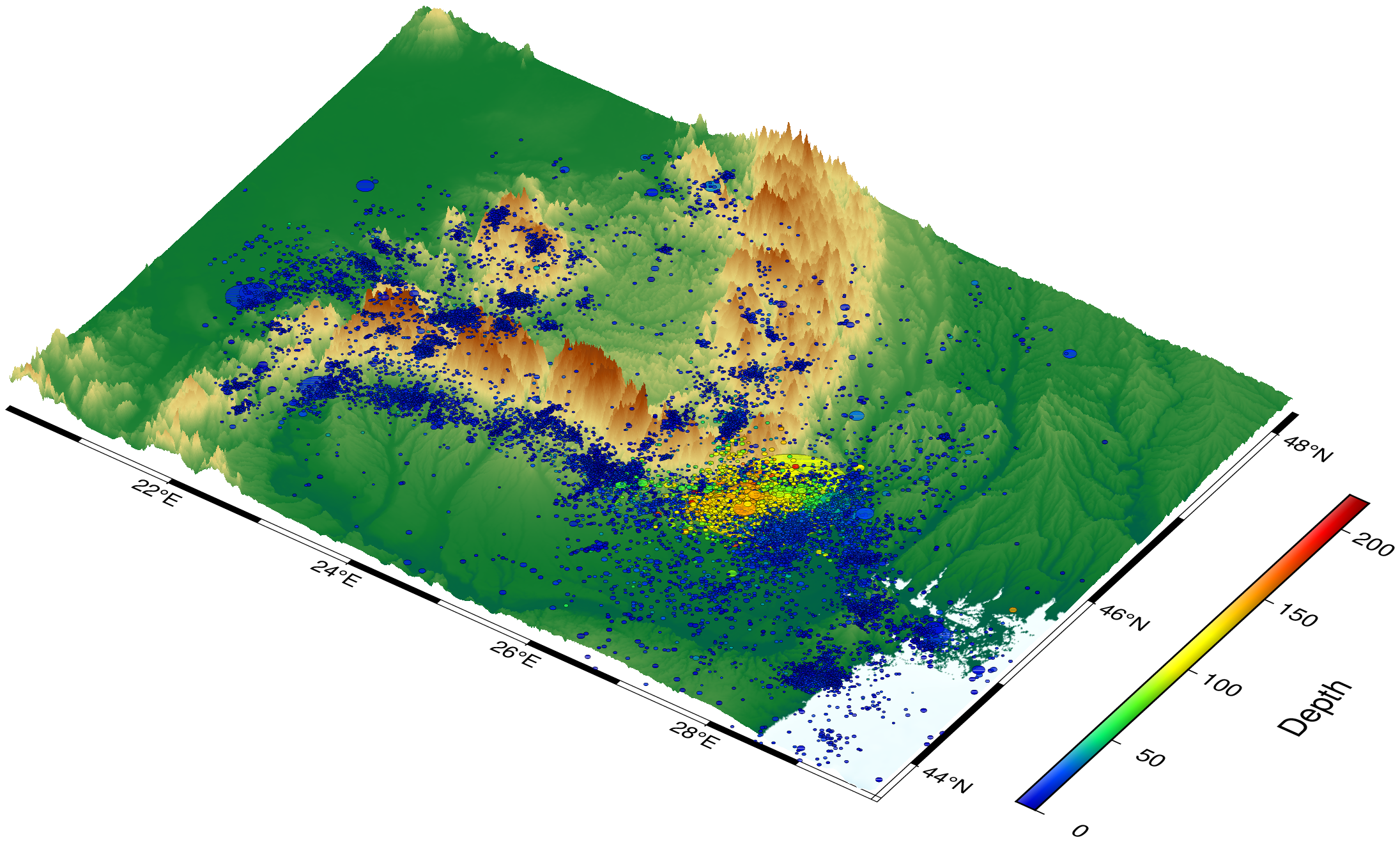}
         \caption{Depth}
         \label{fig:ro_quake_vis_depth}
    \end{subfigure}
    \hfill
    \begin{subfigure}[b]{0.49\textwidth}
         \centering
         \includegraphics[width=\textwidth]{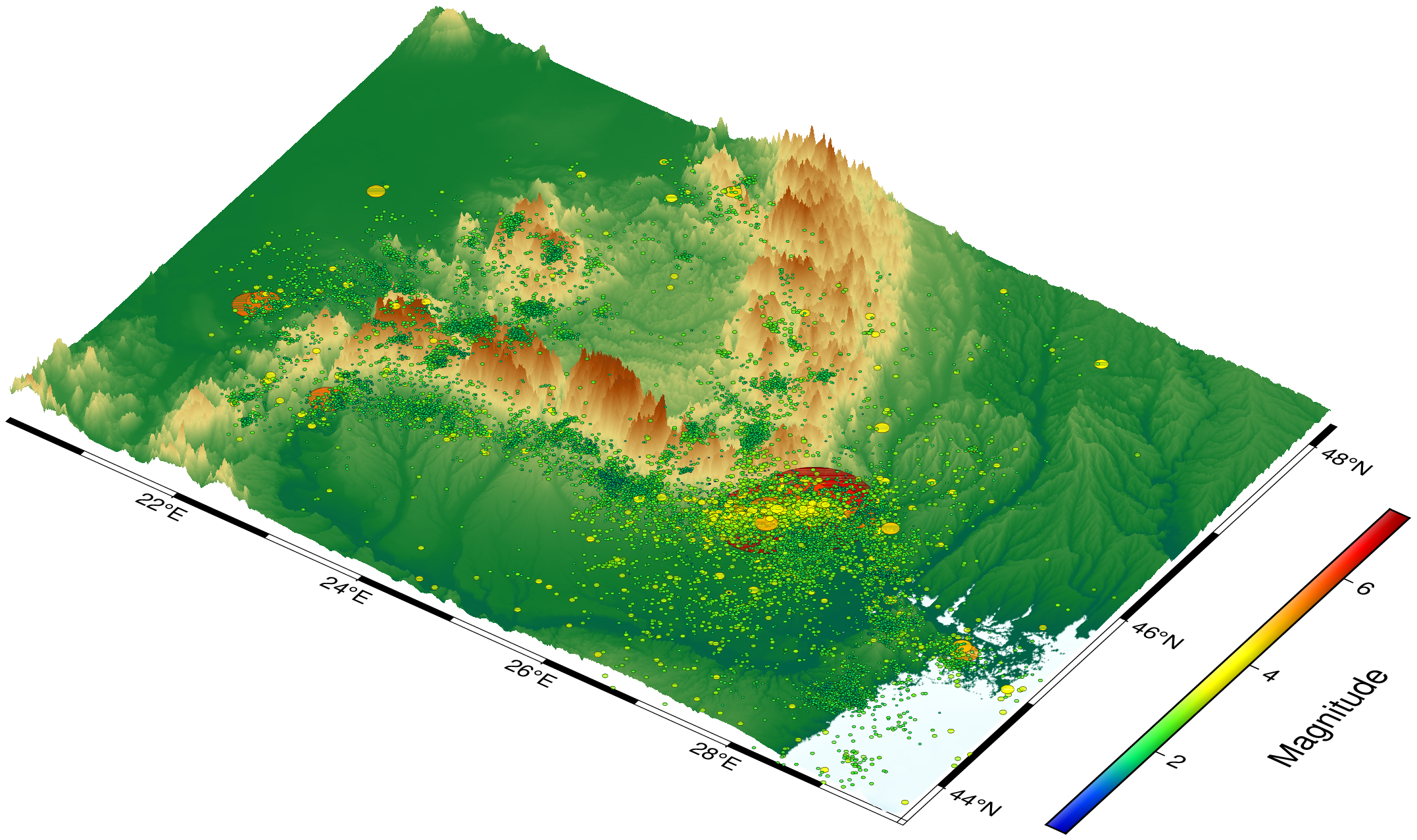}
         \caption{Magnitude}
         \label{fig:ro_quake_vis_mag}
    \end{subfigure}

    \caption{Visualization of earthquakes in Romania with 31378 events from 1977-03-04 until 2022-12-29 with magnitudes ranging between $0.1$ and $7.4$. The color gradient represents \textit{(left)} the depth and \textit{(right)} the magnitude of the earthquakes.}
    \label{fig:ro_quake_vis}
\end{figure*}

\begin{figure*}[hbt!]
    \centering
    \begin{subfigure}[b]{0.49\textwidth}
         \centering
         \includegraphics[width=\textwidth]{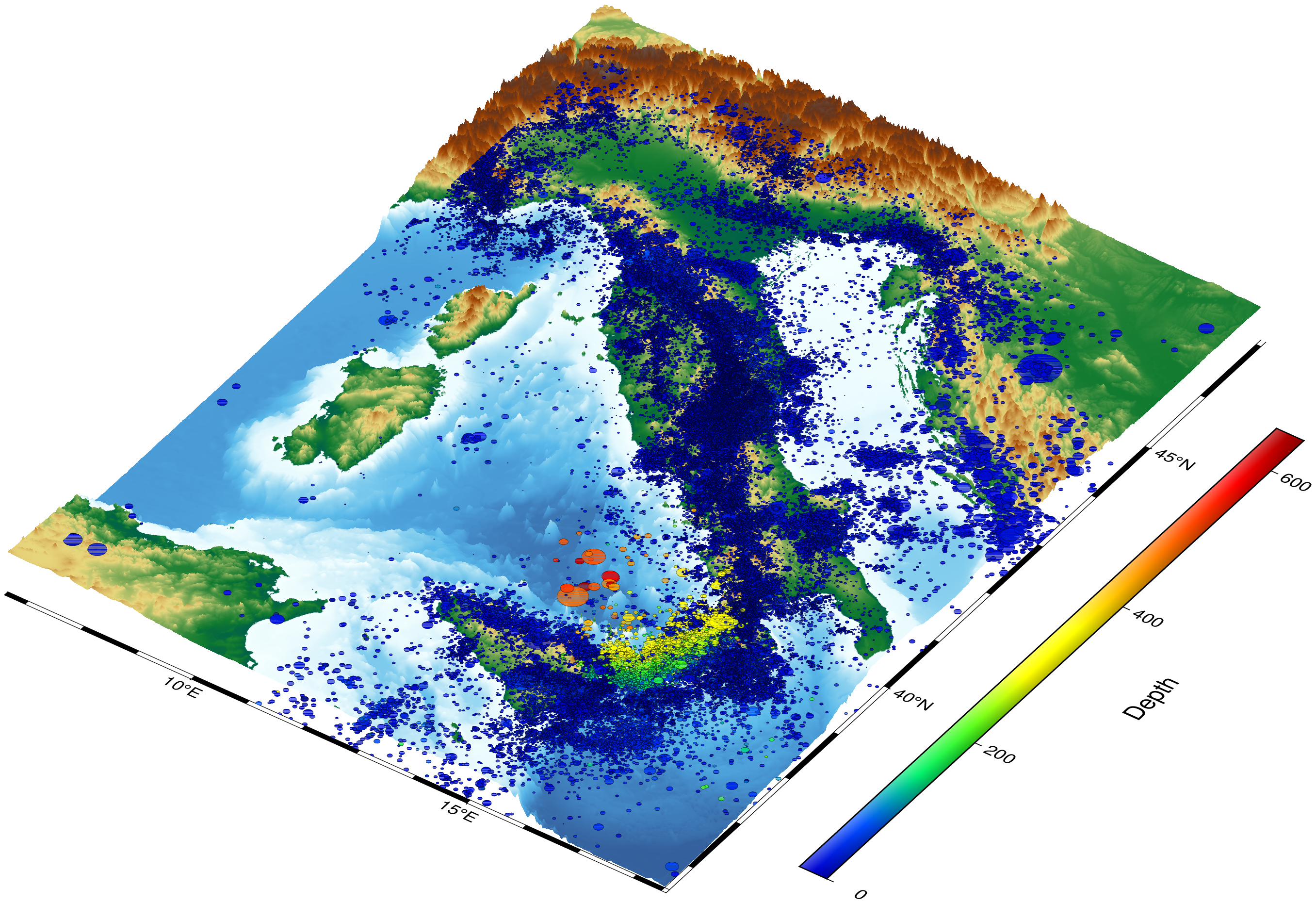}
         \caption{Depth}
         \label{fig:it_quake_vis_depth}
    \end{subfigure}
    \hfill
    \begin{subfigure}[b]{0.49\textwidth}
         \centering
         \includegraphics[width=\textwidth]{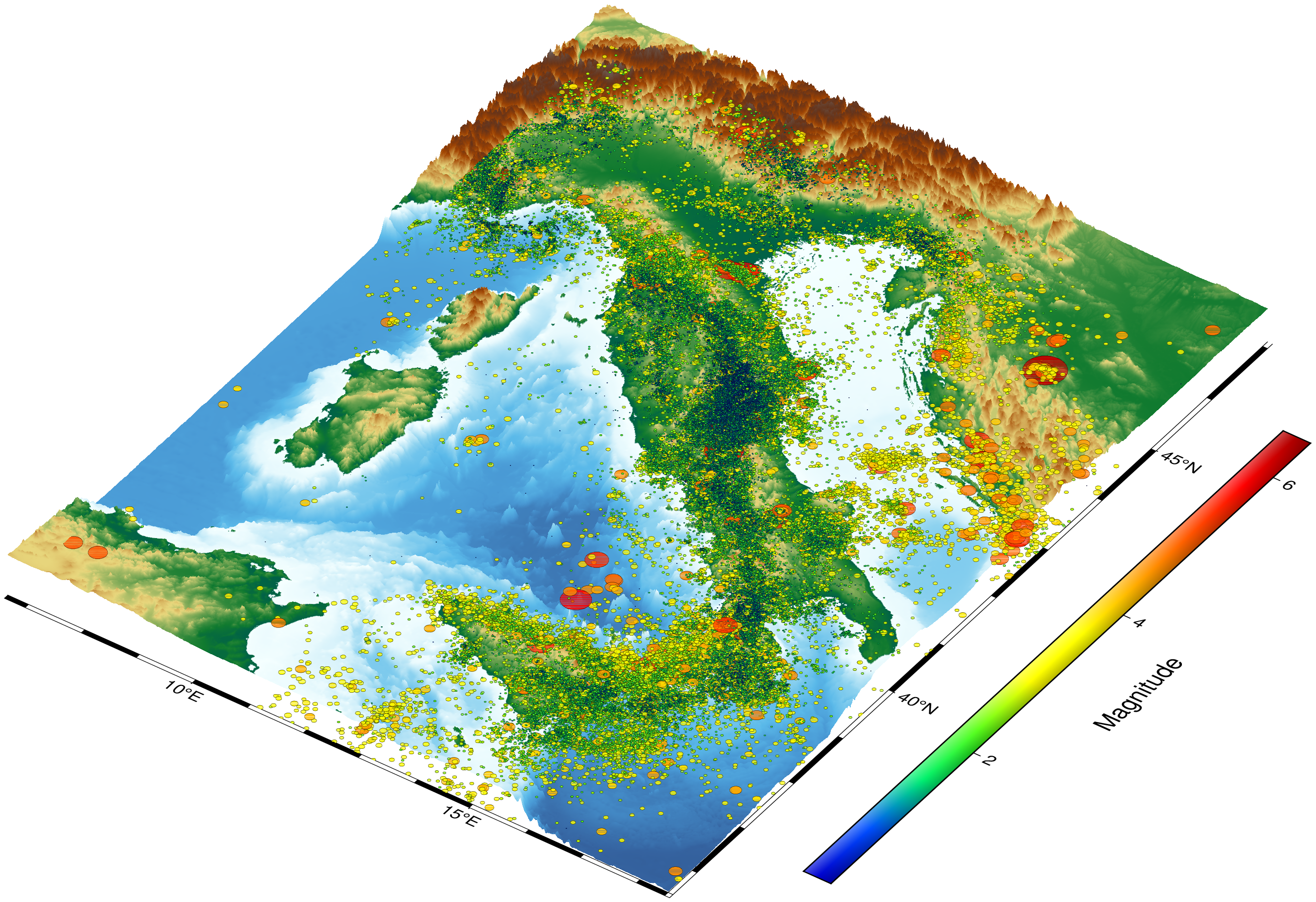}
         \caption{Magnitude}
         \label{fig:it_quake_vis_mag}
    \end{subfigure}

    \caption{Visualization of earthquakes in Italy with 410234 events from 1985-01-02 until 2023-03-08 with magnitudes ranging between $0.1$ and $6.5$.}
    \label{fig:it_quake_vis}
\end{figure*}

\begin{figure*}[hbt!]
    \centering
    \begin{subfigure}[b]{0.49\textwidth}
         \centering
         \includegraphics[width=\textwidth]{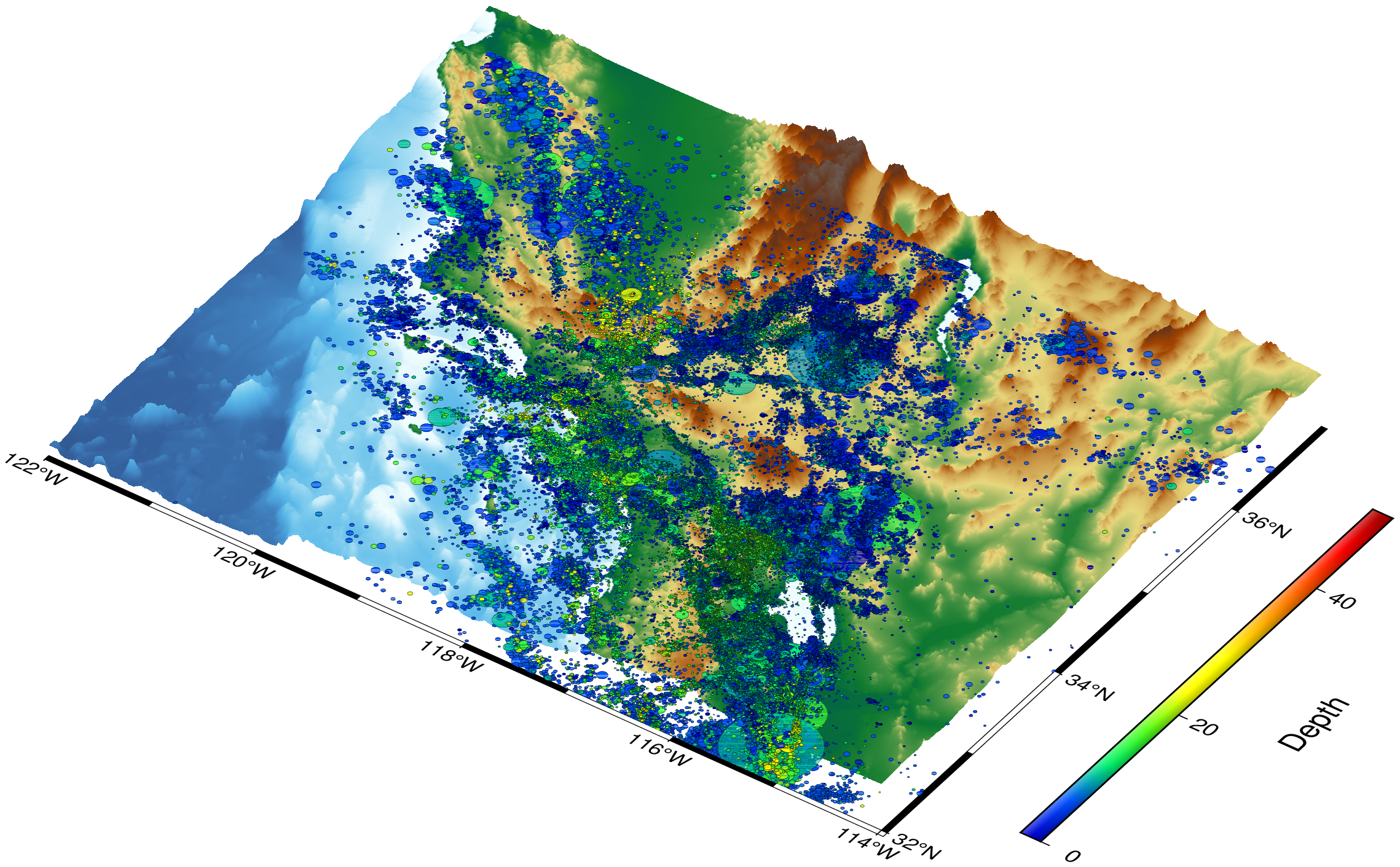}
         \caption{Depth}
         \label{fig:cali_quake_vis_depth}
    \end{subfigure}
    \hfill
    \begin{subfigure}[b]{0.49\textwidth}
         \centering
         \includegraphics[width=\textwidth]{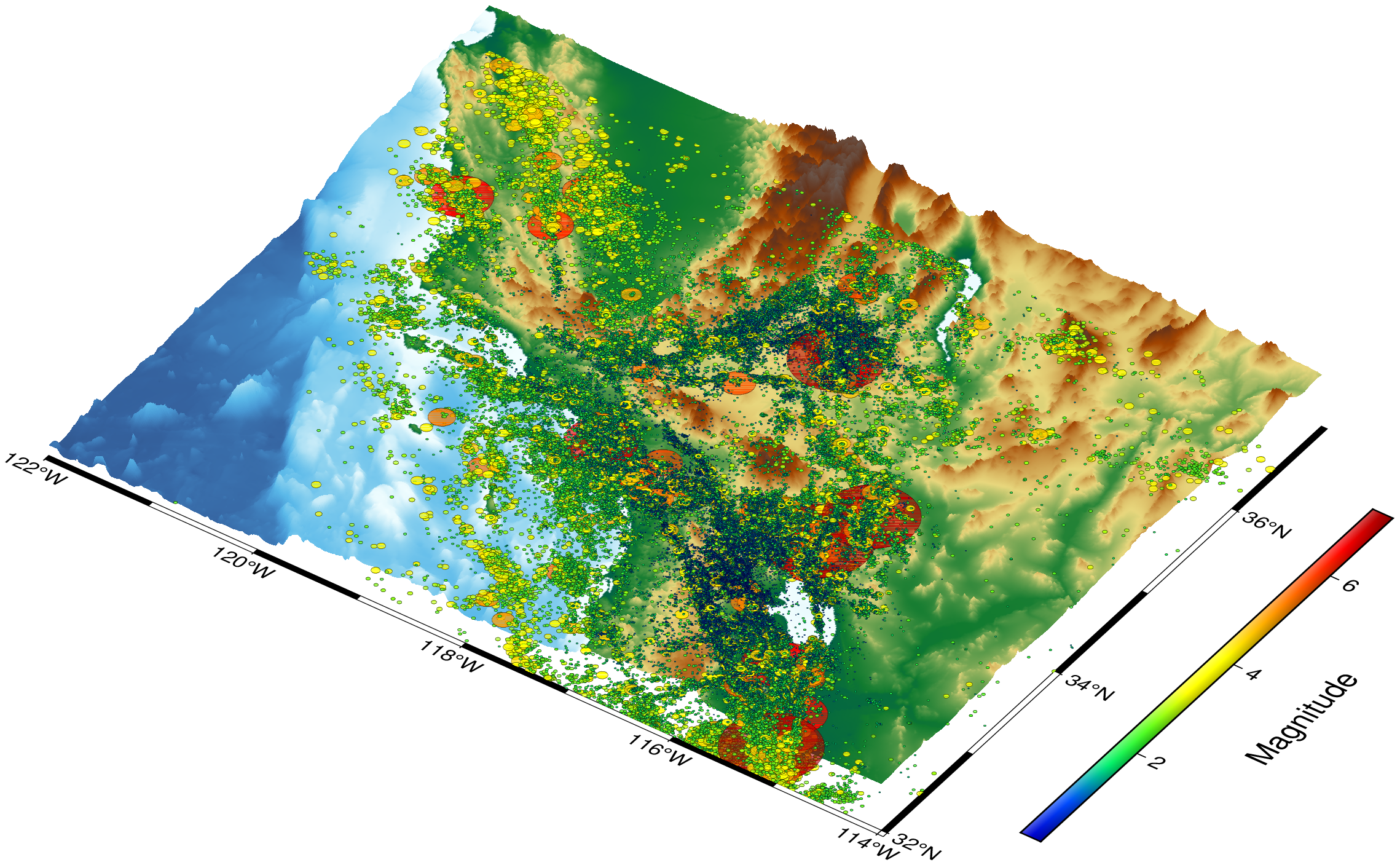}
         \caption{Magnitude}
         \label{fig:cali_quake_vis_mag}
    \end{subfigure}

    \caption{Visualization of earthquakes from the California region with $784754$ events from 01-01-1977 until 07-03-2023 with magnitudes ranging between $0.01$ and $7.3$.}
    \label{fig:cali_quake_vis}
\end{figure*}

\begin{figure*}[hbt!]
    \centering
    \begin{subfigure}[b]{0.49\textwidth}
         \centering
         \includegraphics[width=\textwidth]{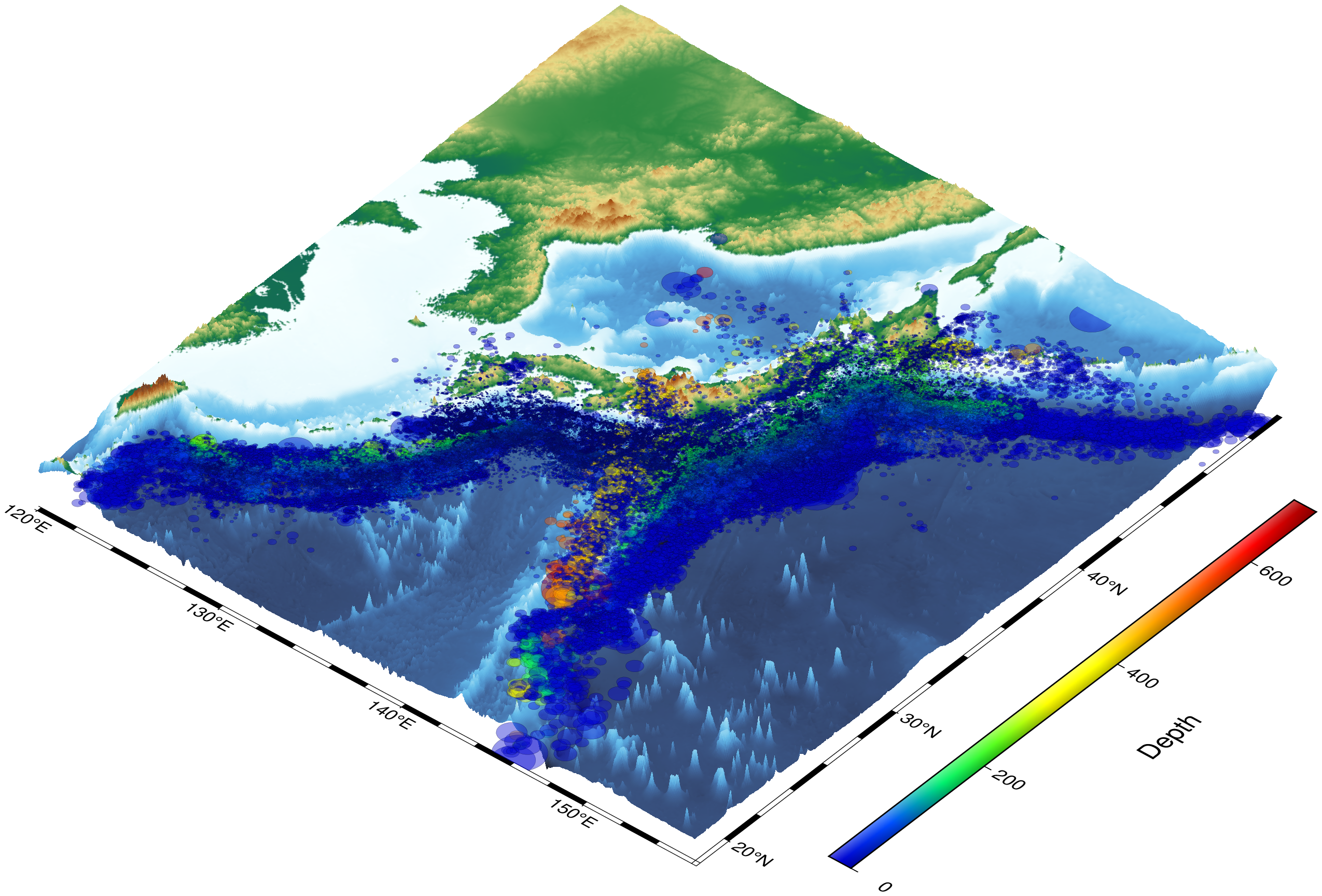}
         \caption{Depth}
         \label{fig:jap_quake_vis_depth}
    \end{subfigure}
    \hfill
    \begin{subfigure}[b]{0.49\textwidth}
         \centering
         \includegraphics[width=\textwidth]{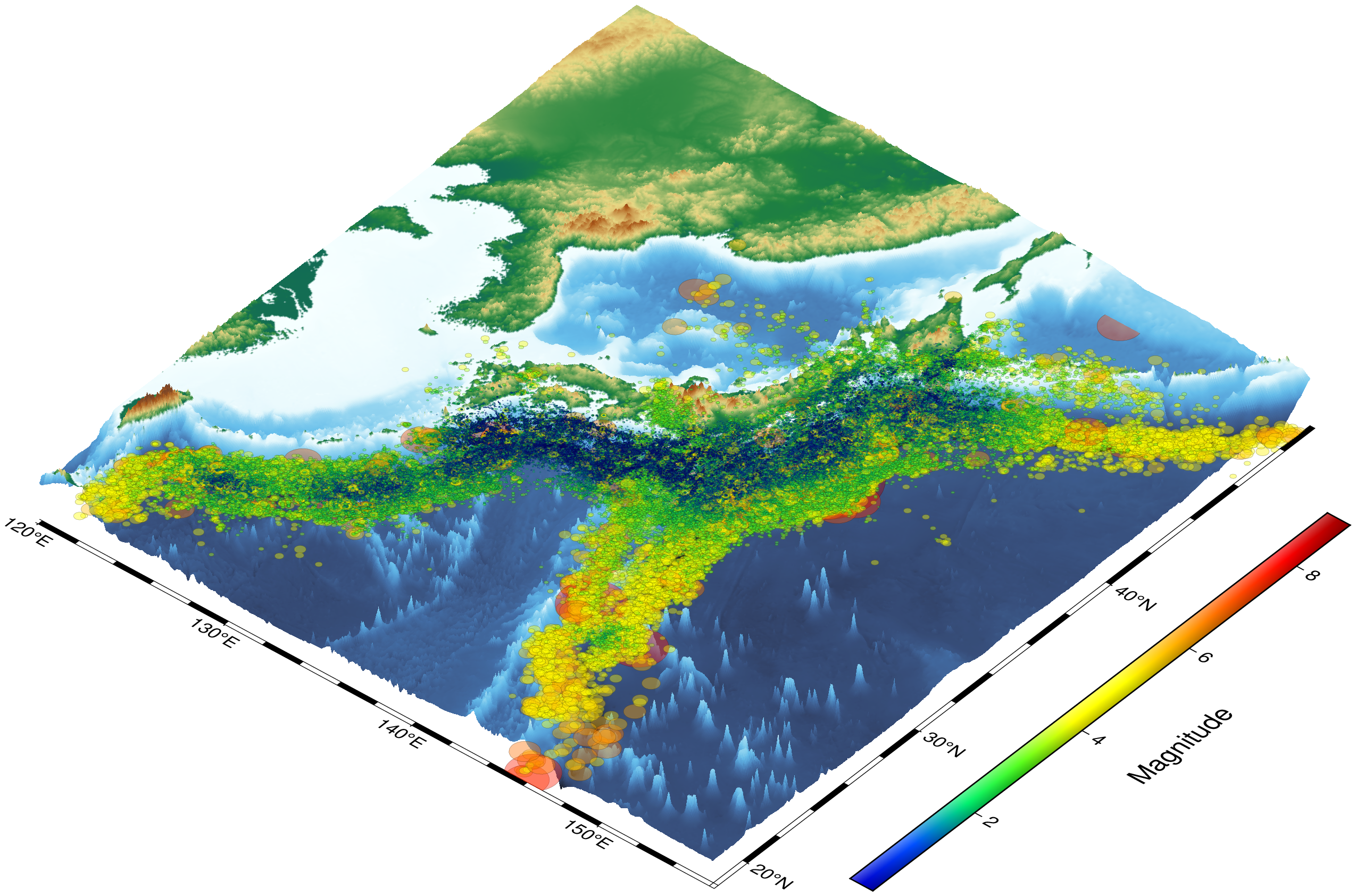}
         \caption{Magnitude}
         \label{fig:jap_quake_vis_mag}
    \end{subfigure}

    \caption{Visualization of a portion of the earthquake database of Japan with 367127 events from 2010-01-01 until 2020-08-31 with magnitudes ranging between $2.0$ and $9.0$.}
    \label{fig:jap_quake_vis}
\end{figure*}

Motivated by our recent findings on scale-free-like distribution of waiting times for earthquakes \cite{vivirschi2020} and seismic events on the Moon and Mars \cite{pana2023}, we investigate here the structure of earthquake networks, using the available earthquake data for Romania, Italy, the California seismic zone in the United States of America, and Japan. We investigate if network motifs have similar or different structures in earthquake networks created for different seismic regions and if there is the prospect of using the scaling exponents to identify seismic regions. Once this approach is extended to even more seismic zones we could potentially infer the correlations between the scaling exponents and seismic mechanism. Complementary to our previous studies where the focus was solely on the magnitude of the observed quakes, we now model the seismic region through a network, following the approach originally introduced by Abe {\it et al.} in Ref. \cite{abe04} and later refined in Refs. \cite{abe2004,abe2006}. This formalism has been used to model different seismic processes, most recently in Refs. \cite{chorozoglou2018, chorozoglou2019, leon2022, martin2022, lotfi2023}. In Refs. \cite{chorozoglou2018, chorozoglou2019} an earthquake network approach is used to study the Greece peninsula, particularly by investigating the small-world property of the networks. In Ref. \cite{min2021} the authors compared seismic activity in the Korean peninsula within two distinct network models, while in Ref. \cite{leon2022} we find a review of the current understanding of earthquake modeling through networks. On a related topic, in Ref. \cite{martin2022} the authors find correlations between the $\alpha$ exponent of the connectivity distribution and the $b$-value from the Gutenberg–Richter law. Lastly, in Ref. \cite{lotfi2023} we find a report on how the time scales used for earthquake networks affect the network's characteristics.

The method proposed by Abe {\it et al.} for constructing earthquake networks differs from that proposed by Baiesi {\it et al.} in Ref. \cite{baiesi2004}. The main difference is that Baiesi {\it et al.} considered a weight associated with the edges of the network to discard weakly linked events, thus identifying correlations between arbitrary pairs of earthquakes, such that aftershocks are better identified. In the article of Abe {\it et al.}, nodes represent the grid cells that the region is split into, and these are linked when subsequent earthquakes occur in them, each edge being treated equally. Within this framework, we show that despite being very different in terms of seismic activity, the aforementioned four seismic regions share striking similarities in terms of network properties. The seismic activity in Romania is almost entirely concentrated in the Vrancea seismic zone, which consists of both crustal and intermediate-depth earthquakes \cite{wenzel1997}, Fig. \ref{fig:ro_quake_vis}. Italy has both surface and very deep earthquakes, present in the Apennine range, which runs from northern to southern Italy and contains several faults running along the entire peninsula, forming a destructive boundary between tectonic plates \cite{pondrelli2020}, Fig. \ref{fig:it_quake_vis}. California presents crustal earthquakes with its main seismic activity along the San Andreas Fault \cite{schulz1993}, Fig. \ref{fig:cali_quake_vis}. Japan has the most intense seismic activity, being at the confluence of four tectonic plates: the Pacific, North American, Eurasian, and Philippines \cite{satake2015}, Fig. \ref{fig:jap_quake_vis}. The databases of the aforementioned seismic zones are publicly available, see \cite{infp, ingv, scedc, jma}, being curated by specialized institutions.

Our article is structured as follows: we report the construction of earthquake networks in Section \ref{sec:earthquake_networks} and detail the fundamental characteristic of a network, the connectivity distribution in \ref{sec:connectivity}. In Section \ref{sec:fitting_power_law}, we briefly discuss previous works \cite{clauset09,alstott14} on the fitting of empirical data which is suspected to have power-law behavior to support our choice of the maximum likelihood estimation method \cite{Bauke2007}. Following this, in Section \ref{sec:goodness_of_fit}, we discuss the use of the Kolmogorov-Smirnov statistic  \cite{Massey1951} to quantify the quality of the fits. Network motifs \cite{latora2017} are introduced in Section \ref{sec:structure_of_motifs}, where we show how they can be used to characterize seismic regions. In Section \ref{sec:computational_framework}, we present the toolbox developed for the analyses of seismic data. The results of our analyses on network connectivity are presented in Section \ref{sec:connectivity_analysis}, and the distributions of three- and four-event motifs in Section \ref{sec:motifs_analysis}. Finally, in Section \ref{sec:conclusion}, we present our conclusions and an outlook on future research. 

All the figures and visualizations are developed with original codes, in an open-source environment using \texttt{Julia} and \texttt{Python} programming languages and are available and documented on GitHub.\footnote{The earthquake networks toolbox is publicly available at \url{https://github.com/gabipana7/seismic-networks}}

\section{\label{sec:method}Method}

\subsection{\label{sec:earthquake_networks}Earthquake Networks}

Earthquake networks are constructed as follows: the considered region is divided into cubic cells in which earthquakes occur, each cube representing a vertex of the network. Two successive events define an edge between two vertices. Successive events occurring in the same cube form a loop. The network is constructed by adding vertices and edges as they result from the earthquake database. In what follows we will refer to the \textit{cell size}, or \textit{cell length}, or simply $L$ as the length in kilometers (km) of a side of one small cube. This number, therefore, defines a network. For example, a network of $L=5\,\mathrm{km}$, means that the seismic region has been split into cubes of $5\, \mathrm{km} \times 5\, \mathrm{km} \times 5\, \mathrm{km}$.

\subsubsection{\label{sec:connectivity}Connectivity}

One of the fundamental properties of a network is its node connectivity (or degree distribution), and we study it for the earthquake networks in Romania, Italy, California (USA), and Japan. As scale-free networks seem to be ubiquitous, the list of examples of such networks including the World Wide Web, citation networks, movie actor collaboration networks, cellular networks, etc. \cite{albert2002}, we investigate here if the connectivity of earthquake networks is also distributed as a power law. We look at earthquake networks as static objects and disregard their growth, even though one could potentially check the validity of the Barab\'asi model of preferential attachment \cite{barabasi1999} for the evolution of these networks using the available seismic data.  

For the purpose of our study, a scale-free network is one that has a distribution of node connectivity of the form \begin{equation}
    \label{eq:prop_con}
    P_k \sim k^{-\alpha}
\end{equation}
with $\alpha>1$. the value of $\alpha$ ranges between $1$ and $3$ in the empirical examples described in Ref. \cite{albert2002}.

\subsubsection{\label{sec:fitting_power_law}Fitting power-law distributed data}

The proper way of fitting data which is suspected to have a power-law distribution has been a topic of discussion in the literature \cite{clauset09,alstott14}. From linear binning to logarithmic binning, it is shown also in Ref. \cite{Virkar2014} that binning can lead to inaccurate data-fitting results. Due to noise in the tail and the reduction of data by binning, fitting the slope on a logarithmic plot is known to introduce systematic biases to the value of the exponent. An alternative, more accurate method relies on using cumulative distribution functions and computing the $\alpha$ exponents via maximum likelihood estimation method.

A cumulative distribution function (CDF) of a real-valued random variable $X$ is given by
\begin{equation}
    F_X(x) = P(X \leq x),
\end{equation}
the right side of the equation representing the probability that the variable $X$ takes on a value \textit{less than or equal to} $x$.

In the case of data suspected to follow a power-law distribution, it is useful to study the opposite phenomenon, representing the data via a complementary cumulative distribution function (CCDF)
\begin{equation}
    \hat{F}_X(x)=P(X>x)=1-F_X(x), 
\end{equation}
where the probability that the variable $X$ \textit{strictly exceeds} the value $x$ is depicted by the right side.

For a power-law PDF
\begin{equation}
    p(x) = C x^{-\alpha},
\end{equation}
where $C$ is the normalization constant, the probability $P(x)$ that the variable has a value greater than $x$ is
\begin{equation}
    P(x)=\int_x^\infty C(x')^{-\alpha}dx' = \frac{C}{\alpha-1}x^{-(\alpha-1)}. 
\end{equation}

The preferred method for determining the $\alpha$ exponent is the maximum likelihood estimation (MLE). For MLE we follow the numerical recipe provided in Refs. \cite{clauset09,alstott14} and use the \texttt{Python} \texttt{powerlaw} package presented in Ref. \cite{alstott14}. The exponent is found through the formula
\begin{align}
    &\alpha = 1 + n \left( \sum_{i=1}^n \ln\frac{x_i}{x_{\mathrm{min}}} \right)^{-1}, \\
    &\sigma=\sqrt{n} \left( \sum_{i=1}^n \ln\frac{x_i}{x_{\mathrm{min}}} \right)^{-1} = \frac{\alpha-1}{\sqrt{n}}, 
\end{align}
where $x_i$ with $i=1,...,n$ are the observed values of $x$ such that $x_i \geq x_{\mathrm{min}}$. The minimum value of $x$, denoted as $x_{\mathrm{min}}$, corresponds to the threshold below which the power-law behavior is observed.

Finding the normalization constant $C$ is done by calculating
\begin{equation}
    \begin{split}
        1&=\int_{x_{\mathrm{min}}}^\infty p(x)dx = C\int_{x_{\mathrm{min}}}^\infty x^{-\alpha} dx \\
        &= \frac{C}{1-\alpha} x^{-\alpha+1}\Big\rvert_{x_{\mathrm{min}}}^\infty,
    \end{split}
\end{equation}
which only holds for $\alpha >1$ since otherwise, the right side of the equation would diverge. When $\alpha>1$ we find that $C=(\alpha-1)x_{\mathrm{min}}^{\alpha-1}$ and the normalized expression for the power law is
\begin{equation}
    p(x) = \frac{\alpha-1}{x_{\mathrm{min}}}\left(\frac{x}{x_{\mathrm{min}}}\right)^{-\alpha}. 
\end{equation}

\subsubsection{\label{sec:goodness_of_fit}Estimating the quality of fit and $x_{\mathrm{min}}$}

There is a variety of measures quantifying the distance between two probability distributions, but for data following non-normal distributions, the most common is the Kolmogorov-Smirnov (KS) statistic \cite{press1992}
\begin{equation}
    D = \underset{x \geq x_{\mathrm{min}}}{\max} |S(x)-P(x)|,
\end{equation}
which represents the maximum distance between the CDFs of the data and the fitted model. Here $S(x)$ is the CDF of the data for observations with values greater than or equal to $x_{\mathrm{min}}$ and $P(X)$ is the CDF for the power-law model that best fits the data in the region $x \geq x_{\mathrm{min}}$. Our estimate of the $x_{\mathrm{min}}$ will be the one that minimizes the KS statistic.

As a first step in our analysis, we calculate how two parameters of the connectivity distribution, namely the $\alpha$ exponent and $x_{\mathrm{min}}$, vary with the cell size across different networks. To this end, we sweep numerically the $L$-interval from $0.5$ to $20$ in increments of $0.5$. For each value of $L$, we calculate $\alpha$ and $x_{\mathrm{min}}$ using the Maximum Likelihood Estimation (MLE), corresponding to each degree distribution. These results are then plotted against the corresponding $L$ values, and the color gradient on the plot represents the quality of fit computed using the Kolmogorov-Smirnov distance. These plots are essential for our subsequent analyses as they reveal which cell sizes correspond to networks having scale-free like degree distributions. We then select a few cell lengths for each of the four seismic regions in focus to construct networks and determine the relevant distributions, namely the connectivity distributions and the distributions of three- and four-event motifs.


\begin{table*}[!htbp]
\begin{ruledtabular}
\begin{tabular}{ccccccc}
            & Events    & Timespan                   & Latitude               & Longitude                          & Depth (km)                     & Magnitude  \\
\hline
Romania     & $31378$   & 1977-03-04 --- 2022-12-29  & $43.59$ --- $48.23$    & $20.19$ --- $29.84$               & $0.0$ --- $218.4$           & $0.1$\phantom{0} --- $7.4$  \\
Italy       & $410234$  & 1985-01-02 --- 2023-03-08  & $35.50$ --- $47.08$    & \phantom{9}$6.61$ ---  $18.51$    & $0.0$ --- $644.4$           & $0.1$\phantom{0} --- $6.5$  \\
California  & $784754$  & 1977-01-02 --- 2023-03-07  & $32.00$ --- $37.00$    & $-122.0$ --- $-114.0$             & $0.0$ --- \phantom{9}$51.1$ & $0.01$ --- $7.3$ \\
Japan       & $3422706$ & 1995-01-01 --- 2020-08-31  & $17.40$ --- $50.42$    & $118.90$ --- $156.68$             & $0.0$ --- $698.4$           & $0.1$\phantom{0} --- $9.0$ 
\end{tabular}
\end{ruledtabular}
\caption{\label{tab:data_info}Information regarding the seismic data used for our analysis.}
\end{table*}

\begin{figure*}[!ht]
    \includegraphics[width=0.99\textwidth]{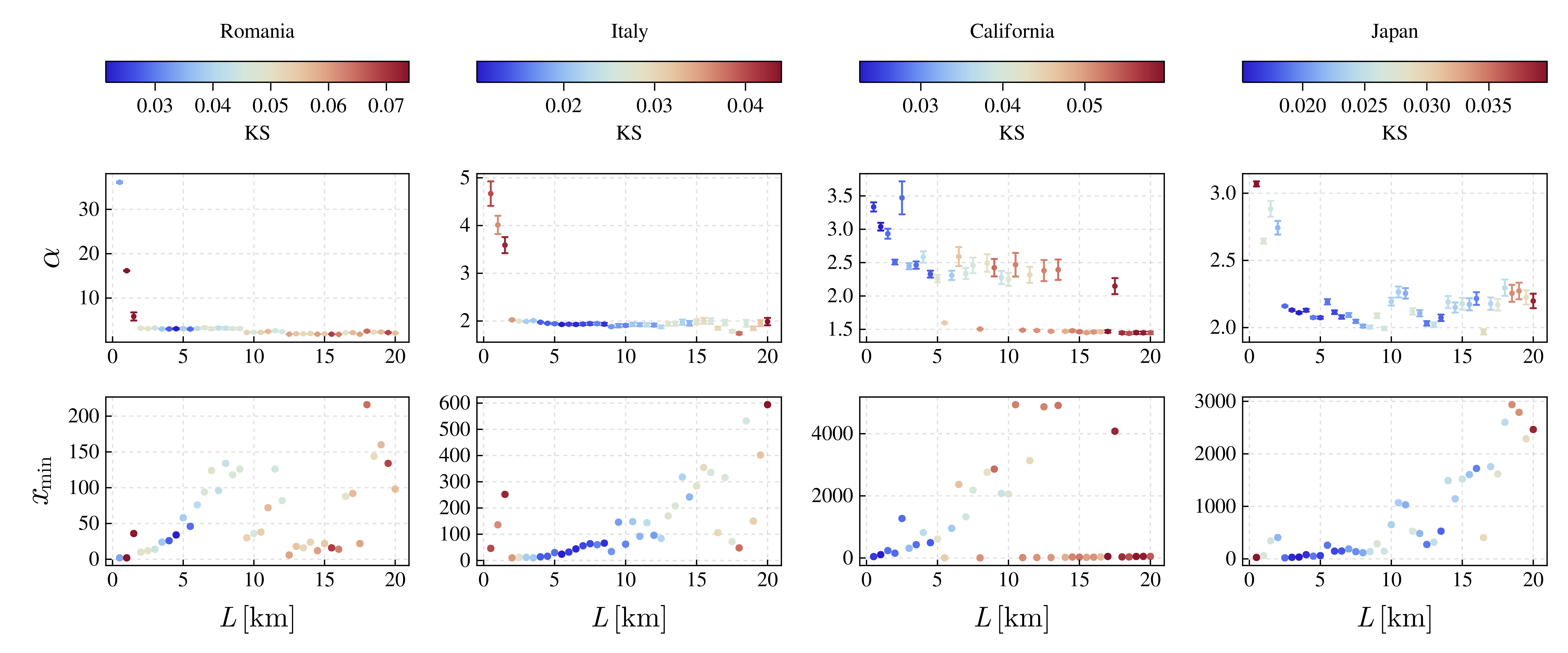}
    \caption{Parameter dependence of $\alpha$ and $x_{\mathrm{min}}$ on the cell size $L$ for networks constructed for Romania, Italy, California, and Japan. The color gradient represents the Kolmogorov-Smirnov distance which measures the fit quality. A low $\mathrm{KS}$ value depicted by the blue color means a better fit, while a high value, depicted by red, means a poor fit.}
    \label{fig:par_dep}
\end{figure*}

\subsection{\label{sec:structure_of_motifs}Structure of motifs}
A network motif is a subgraph or a cycle of length at least three, which is found to occur in real-world complex networks much more frequently than in their corresponding randomized counterparts \cite{latora2017}. Network motifs are specific patterns of local interconnections with potential functional properties, and can be seen as the basic building blocks of real-world networks. For example, motifs of length three, which we refer to as triangles, are highly recurrent in social and biological networks \cite{boccaletti2006}. As another example, let us mention that network motifs are used extensively in molecular biology to study the complexity of protein-protein interaction or genetic regulatory networks \cite{Esti2004}. Motifs are also used in social sciences to uncover hidden interactions between individuals \cite{Wasserman1994}. Drawing from the existing literature on network motifs, we apply the method of motif discovery and analysis to our earthquake networks. In this study, we focus the analysis on three- and four-node motifs, depicting triangles and tetrahedrons in real-world three-dimensional networks. The selected motifs offer significant insight into the structure of earthquake networks, while at the same time being computationally tractable using the current seismic databases. 

Motif analysis is often limited either by the small number of nodes of the network under scrutiny or by the large computing load that comes with motif discovery in big networks. In our study, we have seen both limitations, as the dataset for earthquakes with epicenters in Romania is rather small and the resulting statistic is rather poor, while in the case of Japan obtaining the results was computationally very demanding due to the large number of seismic events in the database. Different tools and methods have been developed to analyze motifs in complex networks, see, for instance, MODA \cite{Omidi2009} and Grochow–Kellis (GK) \cite{Grochow2007}. Currently, one of the best tools for motif discovery is \texttt{Nemomap} \cite{huynh2018}, an improved motif-centric algorithm, which adds upon MODA and GK, being more computationally efficient in mapping complex motif patterns. We use the \texttt{Python} code for \texttt{Nemomap}\footnote{The code for NemoMapPy is publicly available at \url{https://github.com/zicanl/NemoMapPy}} on our earthquake networks to find the three- and four-node motifs. We note that for the motifs with four nodes, we search only for squares and then construct the tetrahedron by adding the corresponding edges.

Since the networks depict real space in the seismic zone, triangles can define an area and tetrahedrons represent a volume. For our analysis, we calculate the areas and volumes of these motifs using each node's position in real space.

For triangles, given the position of nodes, we first calculate the lengths of the three sides ($a$, $b$, and $c$) and then compute the area using Heron's formula
\begin{equation}
    A = \sqrt{s(s-a)(s-b)(s-c)} ,\quad s \equiv \frac{a+b+c}{2}.
\end{equation}

For tetrahedrons, given the position of nodes, we calculate the lengths of the six sides $(U, V, W, u, v, w)$, then use the 3D Heron formula to compute the volume

\begin{equation}
\begin{split}
    V =& \frac{1}{192uvw}\times[(s+r)^2-(p-q)^2]^{1/2} \\
    &\times [(p+q)^2-(r-s)^2]^{1/2}
\end{split}
\end{equation}
where $A \equiv (w-U+v)\cdot(U+v+w), B \equiv (u-V+w)\cdot(V+w+u), C \equiv (v-W+u)\cdot(W+u+v), a \equiv (U-v+w)\cdot(v-w+U), b \equiv (V-w+u)\cdot(w-u+V), c \equiv (W-u+v)\cdot(u-v+W), p \equiv \sqrt{a\cdot B \cdot C}, q \equiv \sqrt{b\cdot C\cdot A}, r \equiv \sqrt{c\cdot A\cdot B}, s \equiv \sqrt{a\cdot b\cdot c}$.

After computing these quantities for each of the motifs identified by \texttt{NemoMap}, we weigh the areas of three-event motifs and volumes of four-event motifs with the total energy released by the earthquakes in each motif. To compute the energy release, we use the magnitudes from existing databases \cite{infp, ingv, scedc, jma}, which have been only reprocessed to fit our codes, keeping the information homogeneous, calculating the energy release using the Gutenberg-Richter formula.

\begin{figure*}[hbt!]
    \centering
    \begin{subfigure}[b]{0.49\textwidth}
         \centering
         \includegraphics[width=\textwidth]{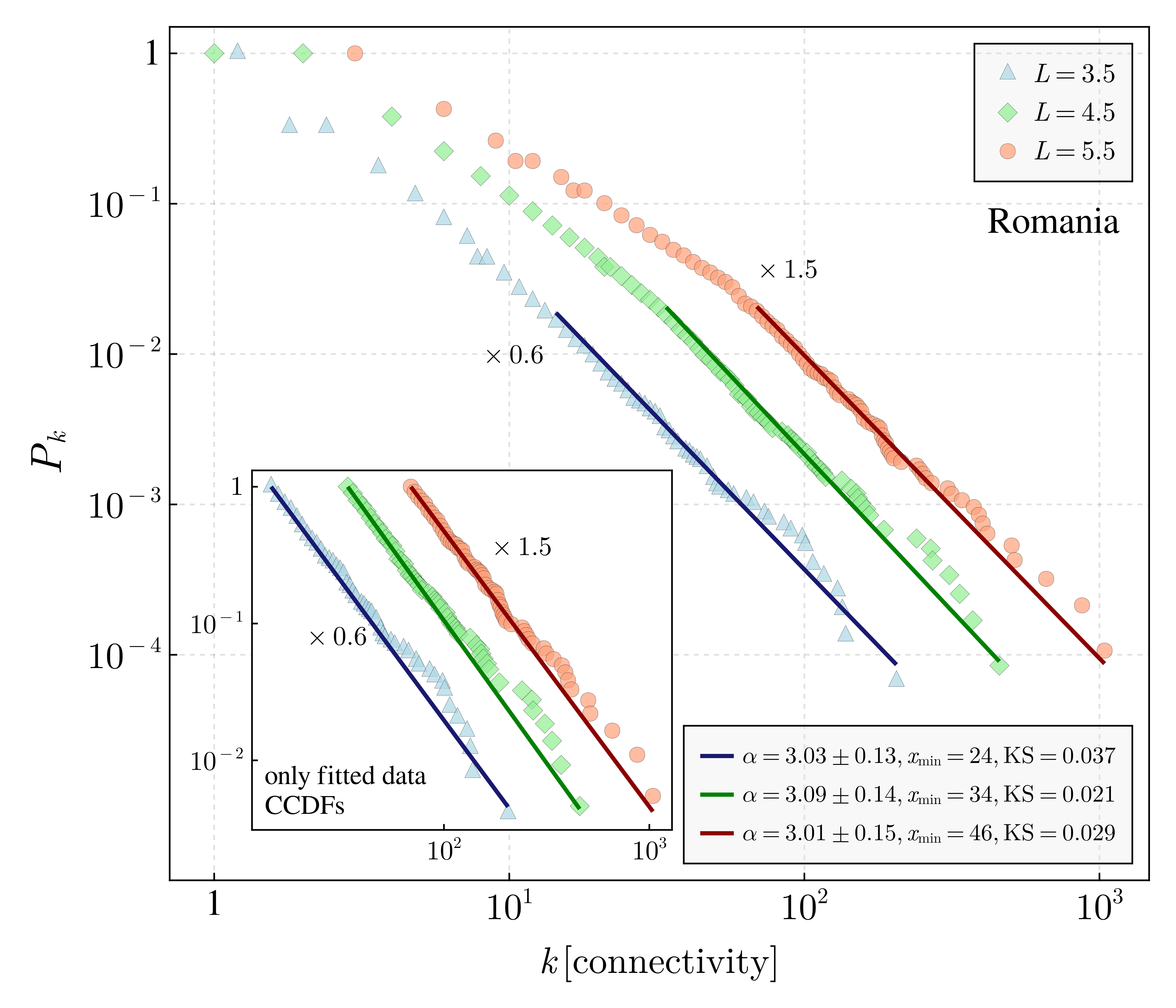}
         \caption{Romania}
         \label{fig:ro_con_best_fits}
    \end{subfigure}
    \hfill
    \centering
    \begin{subfigure}[b]{0.49\textwidth}
         \centering
         \includegraphics[width=\textwidth]{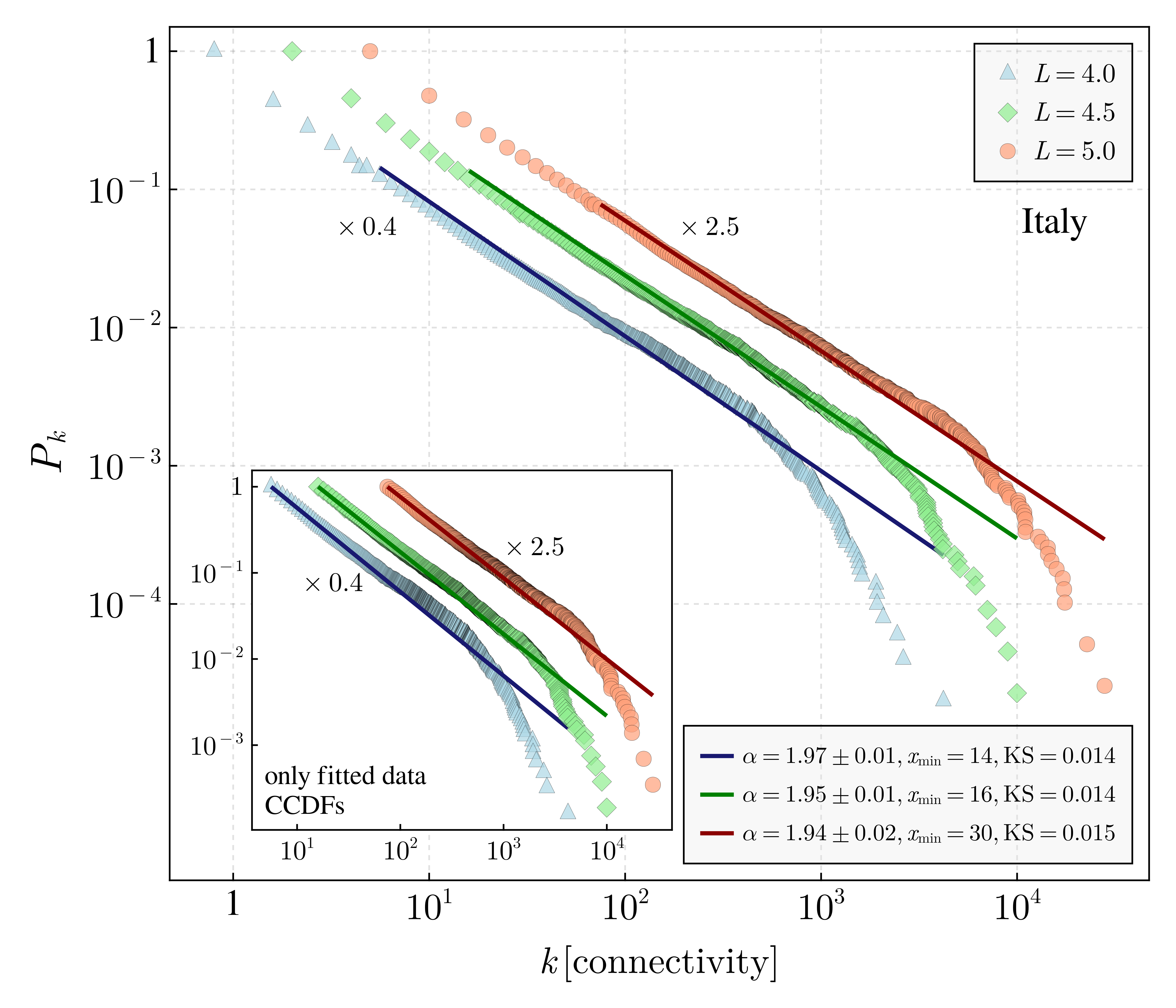}
         \caption{Italy}
         \label{fig:it_con_best_fits}
    \end{subfigure}
    \\
    \centering
    \begin{subfigure}[b]{0.49\textwidth}
         \centering
         \includegraphics[width=\textwidth]{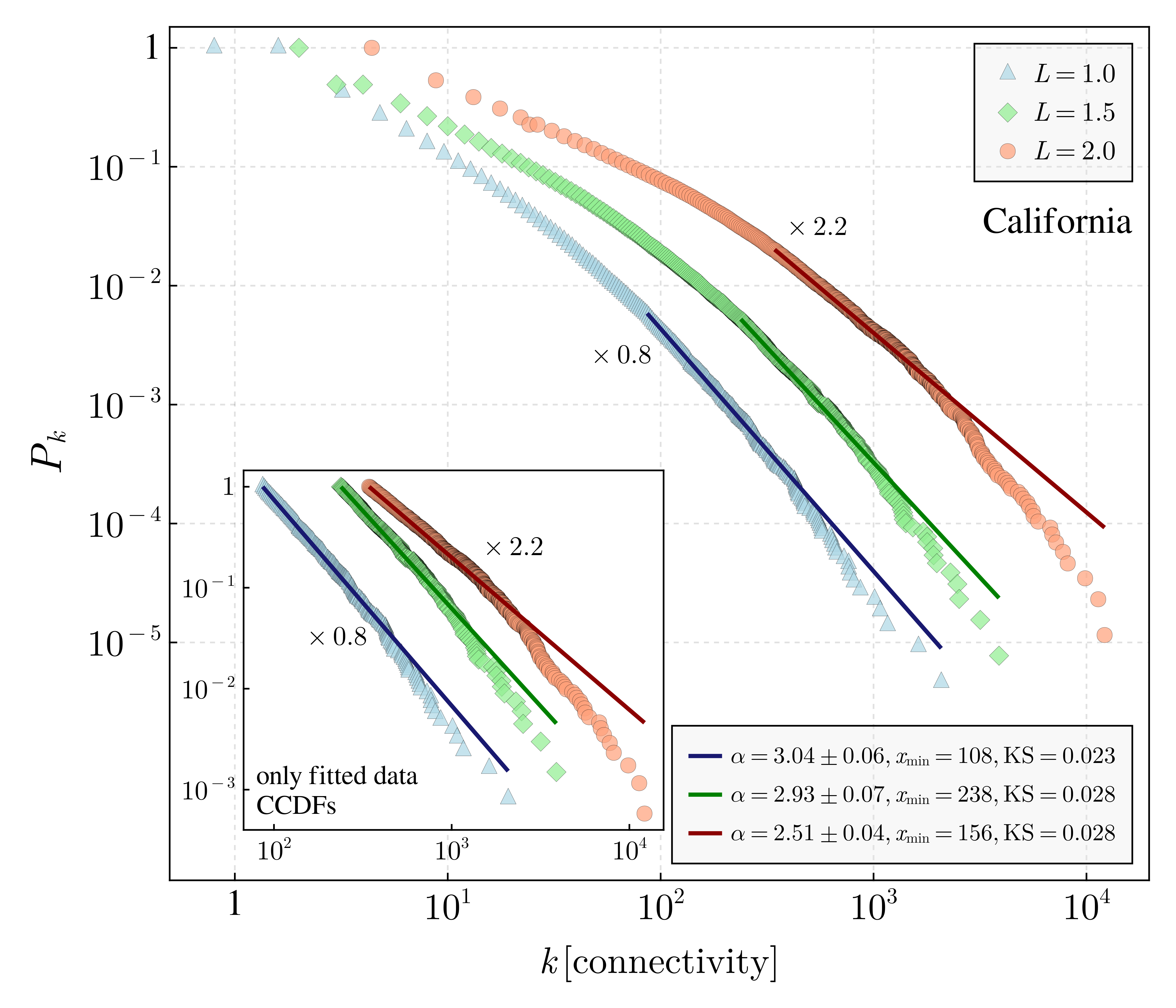}
         \caption{California}
         \label{fig:cali_con_best_fits}
    \end{subfigure}
    \hfill
    \centering
    \begin{subfigure}[b]{0.49\textwidth}
         \centering
         \includegraphics[width=\textwidth]{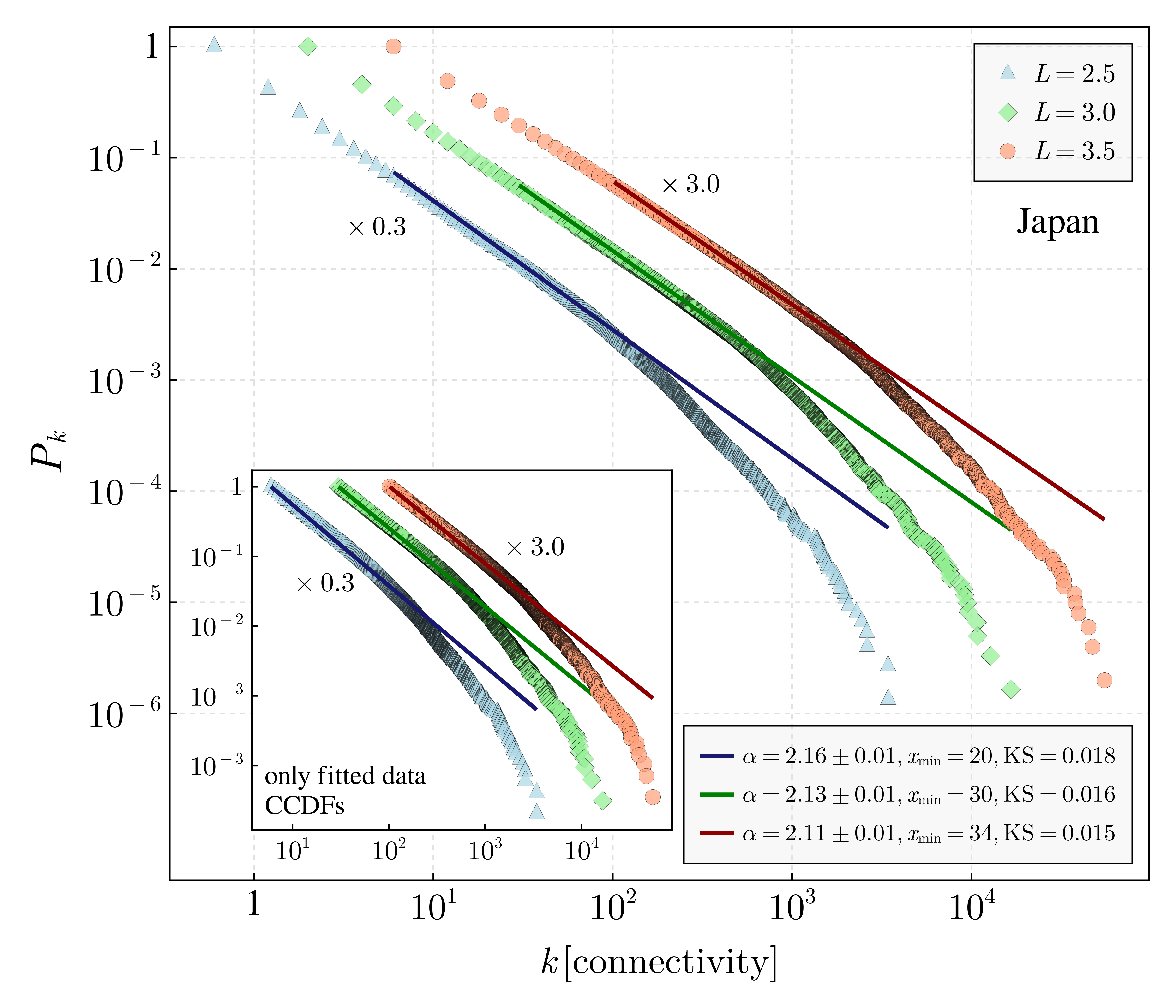}
         \caption{Japan}
         \label{fig:jap_con_best_fits}
    \end{subfigure}

    \caption{Complementary cumulative distribution functions for the connectivity of three networks corresponding to different cube sizes $L$ created with data from (a) Romania, (b) Italy, (c) California, and (d) Japan. The theoretical distribution is represented as a fit through the data, starting with the value $x_{\mathrm{min}}$ that minimizes the Kolmogorov-Smirnov distance. The inset presents only the CCDFs and fits of the data that is considered to behave as a power law, see Eq.~\eqref{eq:prop_con}. Some of the distributions and fits are scaled by a factor depicted on the plot for better visualization.}
    \label{fig:con_best_fits}
\end{figure*}

\subsection{\label{sec:computational_framework}Computational framework}
We have developed an earthquake networks toolbox, \texttt{seismic-networks}, publicly available on GitHub,\footnote{The earthquake networks toolbox is publicly available at \url{https://github.com/gabipana7/seismic-networks}} that contains a series of codes that can be used to automatically analyze the structure of earthquake networks, \textit{i.e.} network construction, connectivity distribution, and distributions of three- and four-event motifs. The codes work with standardized seismic databases, which means that for most analyses, some small curation of the existing datasets is required. The standardized format used by our codes requires the events' timestamp, latitude, longitude, depth, and magnitude. With the help of the codes in the toolbox, the user can generate earthquake networks using different discretization lengths $L$ of the 3D real space, and can compute the connectivity distribution of the network using the \texttt{powerlaw} package to analyze the parameter dependency of $\alpha$ and $x_{\mathrm{min}}$ based on the discretization length $L$. To filter out micro-earthquakes, a magnitude cut-off $M_{\mathrm{min}}=2,3$ is used for the distribution functions of three- and four-event motifs, which are identified using the \texttt{NemoMapPy} \cite{preston2019} (\texttt{Python} version of \texttt{Nemomap}) software. 

For interoperability and open-source disponibility, the toolbox is written in \texttt{Julia} and \texttt{Python}. We applied the codes to the four different regions mentioned in Section \ref{sec:introduction}: Romania, Italy, California (USA), and Japan. For these regions, after collecting the databases and some minor cleaning and trimming, the remaining data is described in Table \ref{tab:data_info}.

\section{\label{sec:results}Results}

\begin{figure*}[hbt!]
    \centering
    \begin{subfigure}[b]{0.49\textwidth}
         \centering
         \includegraphics[width=\textwidth]{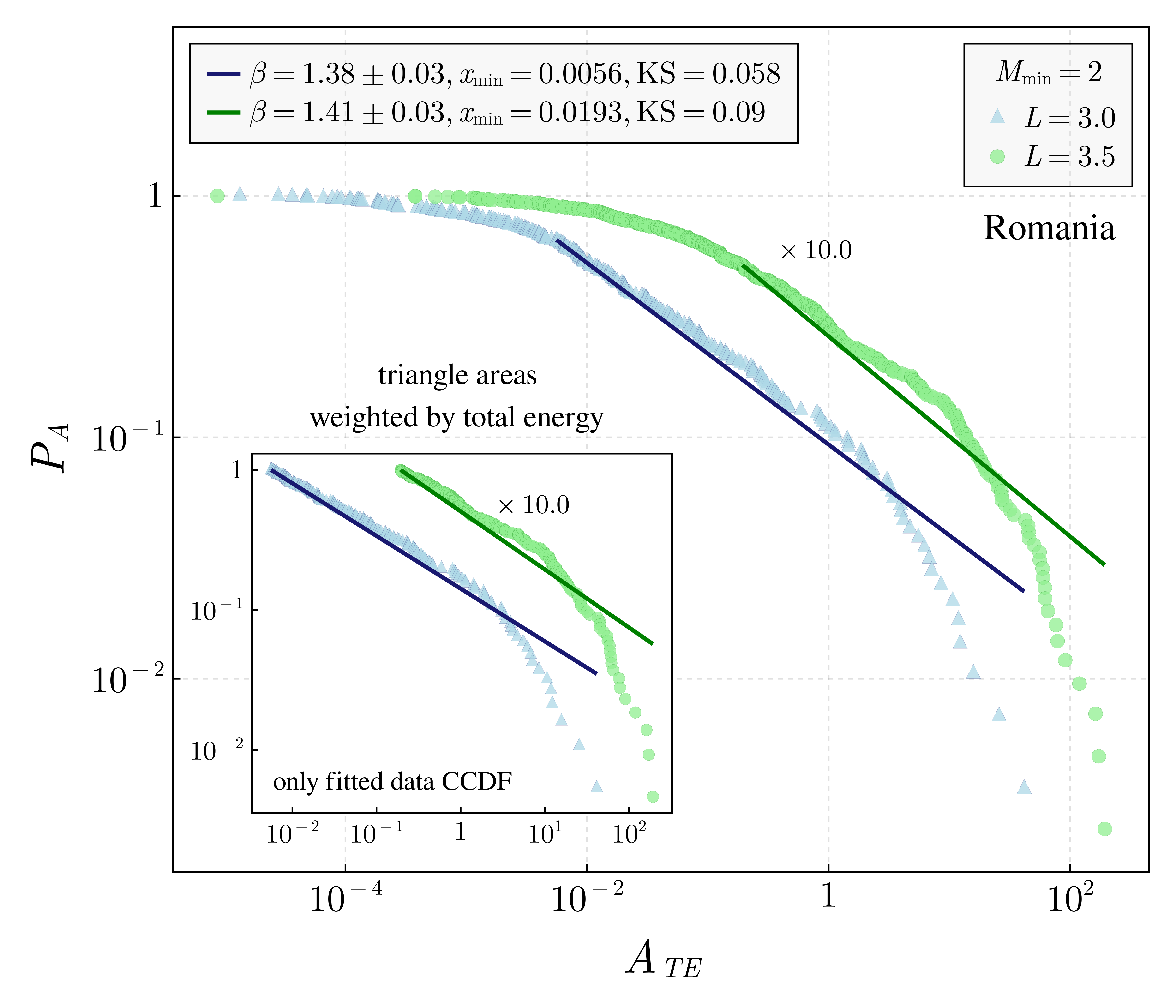}
         \caption{Triangles}
         \label{fig:ro_triangle}
    \end{subfigure}
    \hfill
    \begin{subfigure}[b]{0.49\textwidth}
         \centering
         \includegraphics[width=\textwidth]{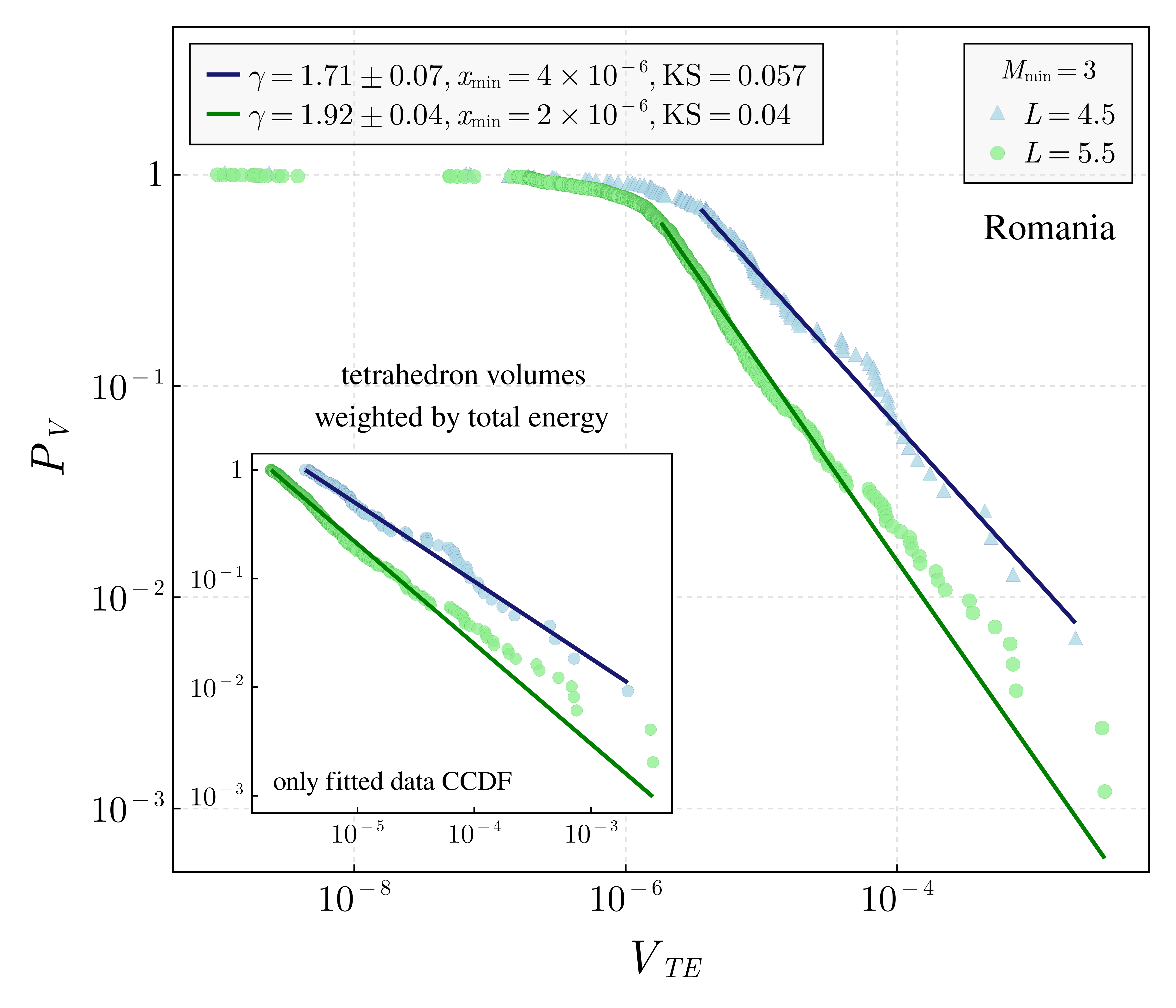}
         \caption{Tetrahedrons}
         \label{fig:ro_tetra}
    \end{subfigure}

    \caption{Complementary cumulative distribution functions for network motifs discovered in networks corresponding to different cube sizes $L$ and magnitude thresholds $M_{\mathrm{min}}$ in Romania. The distribution of triangle areas weighted by the total energy released in motif is present in (a) with an inset presenting only the power-law part of the distribution. The analysis for $L=3.5$ is shifted by a factor of $10$ for better visualization. The distribution of tetrahedron volumes weighted by total energy is present in (b) with an inset similar to the left panel.}
    \label{fig:ro_motifs}
\end{figure*}

\begin{figure*}[hbt!]
    \centering
    \begin{subfigure}[b]{0.49\textwidth}
         \centering
         \includegraphics[width=\textwidth]{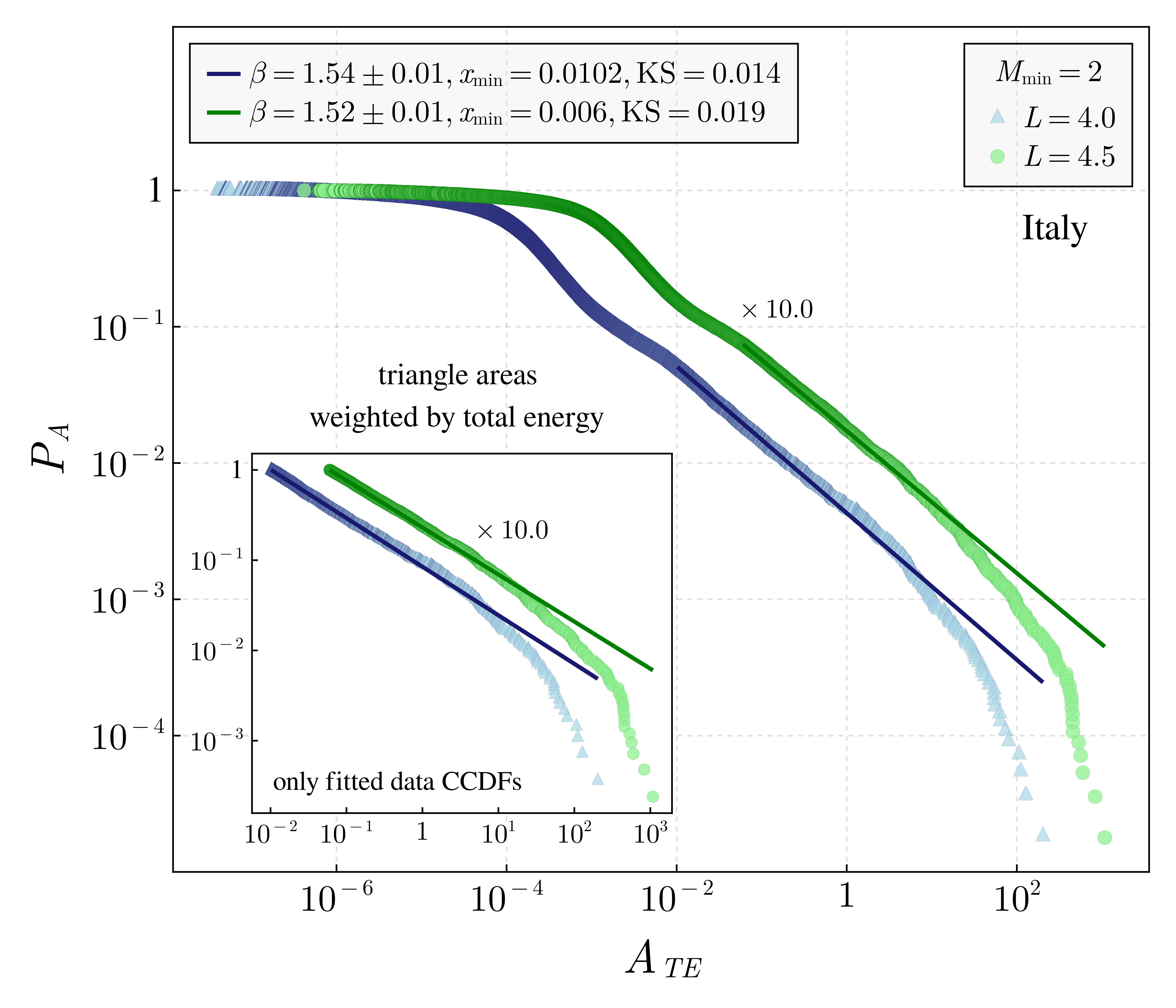}
         \caption{Triangles}
         \label{fig:it_triangle}
    \end{subfigure}
    \hfill
    \begin{subfigure}[b]{0.49\textwidth}
         \centering
         \includegraphics[width=\textwidth]{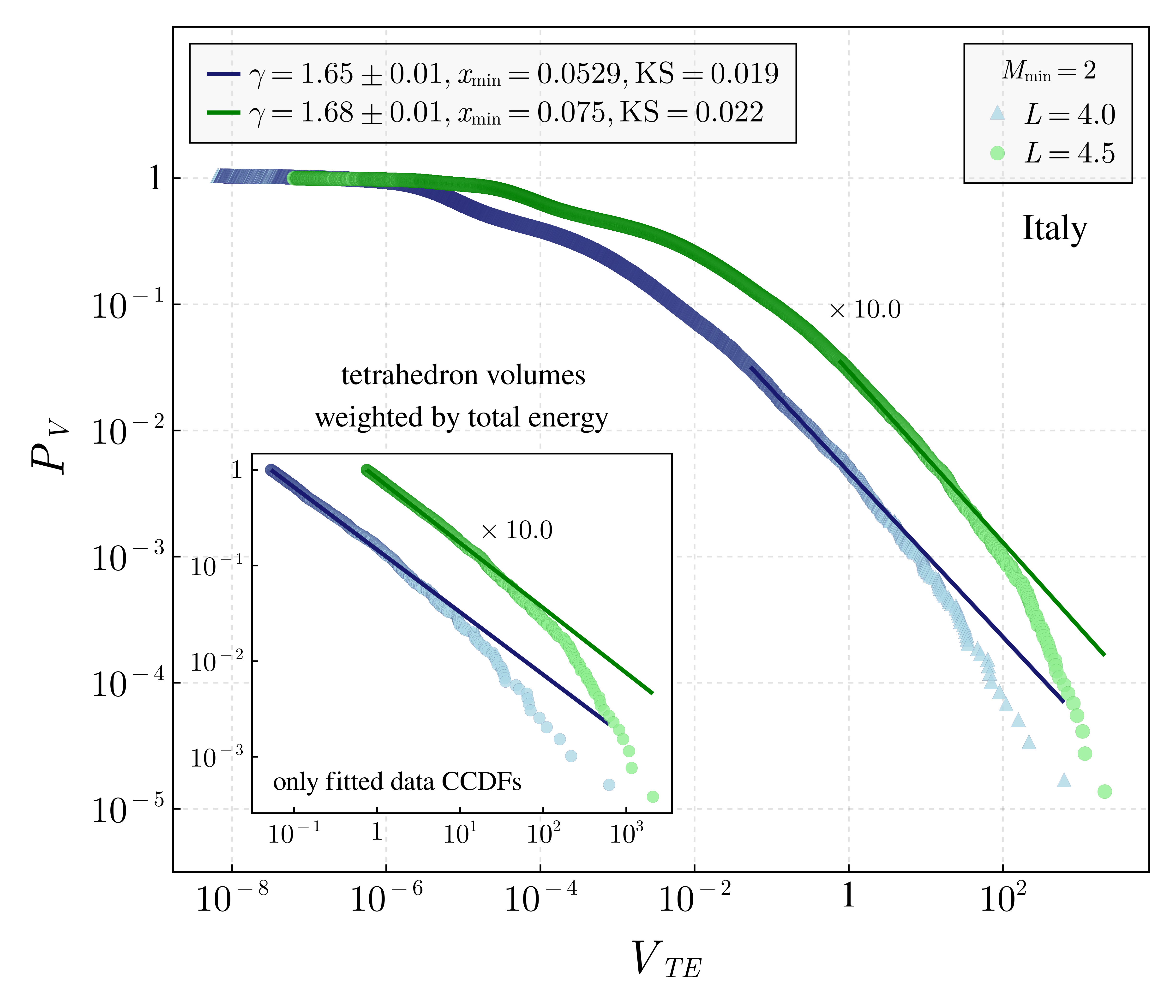}
         \caption{Tetrahedrons}
         \label{fig:it_tetra}
    \end{subfigure}

    \caption{Complementary cumulative distribution functions for network motifs discovered in networks corresponding to different cube sizes $L$ and a magnitude threshold $M_{\mathrm{min}}=2$ in Italy. Details are as in Fig. \ref{fig:ro_motifs}. In both panels, for $L=4.5$ we shifted the distribution with a factor for better visualization.}
    \label{fig:it_motifs}
\end{figure*}

\begin{figure*}[hbt!]
    \centering
    \begin{subfigure}[b]{0.49\textwidth}
         \centering
         \includegraphics[width=\textwidth]{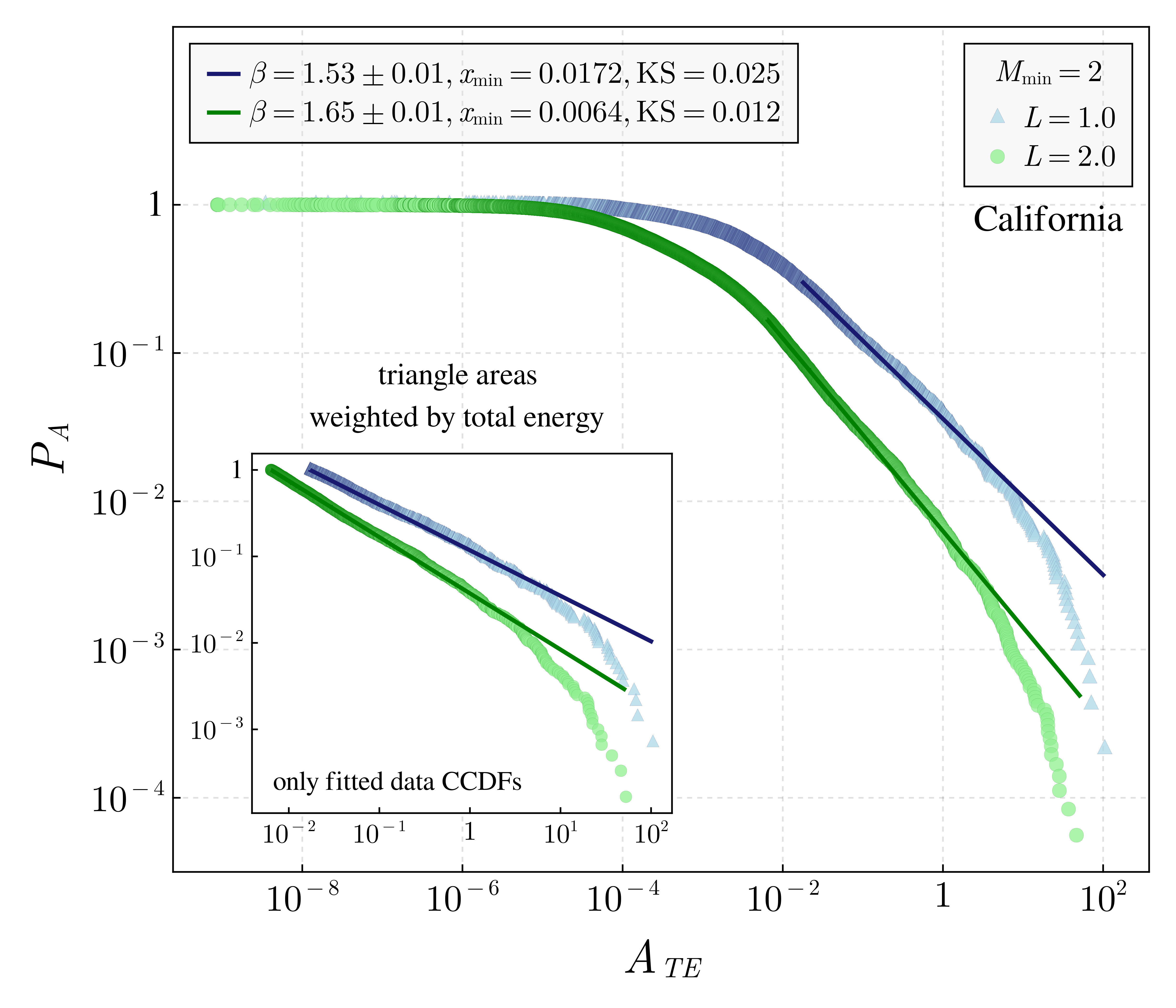}
         \caption{Triangles}
         \label{fig:cali_triangle}
    \end{subfigure}
    \hfill
    \begin{subfigure}[b]{0.49\textwidth}
         \centering
         \includegraphics[width=\textwidth]{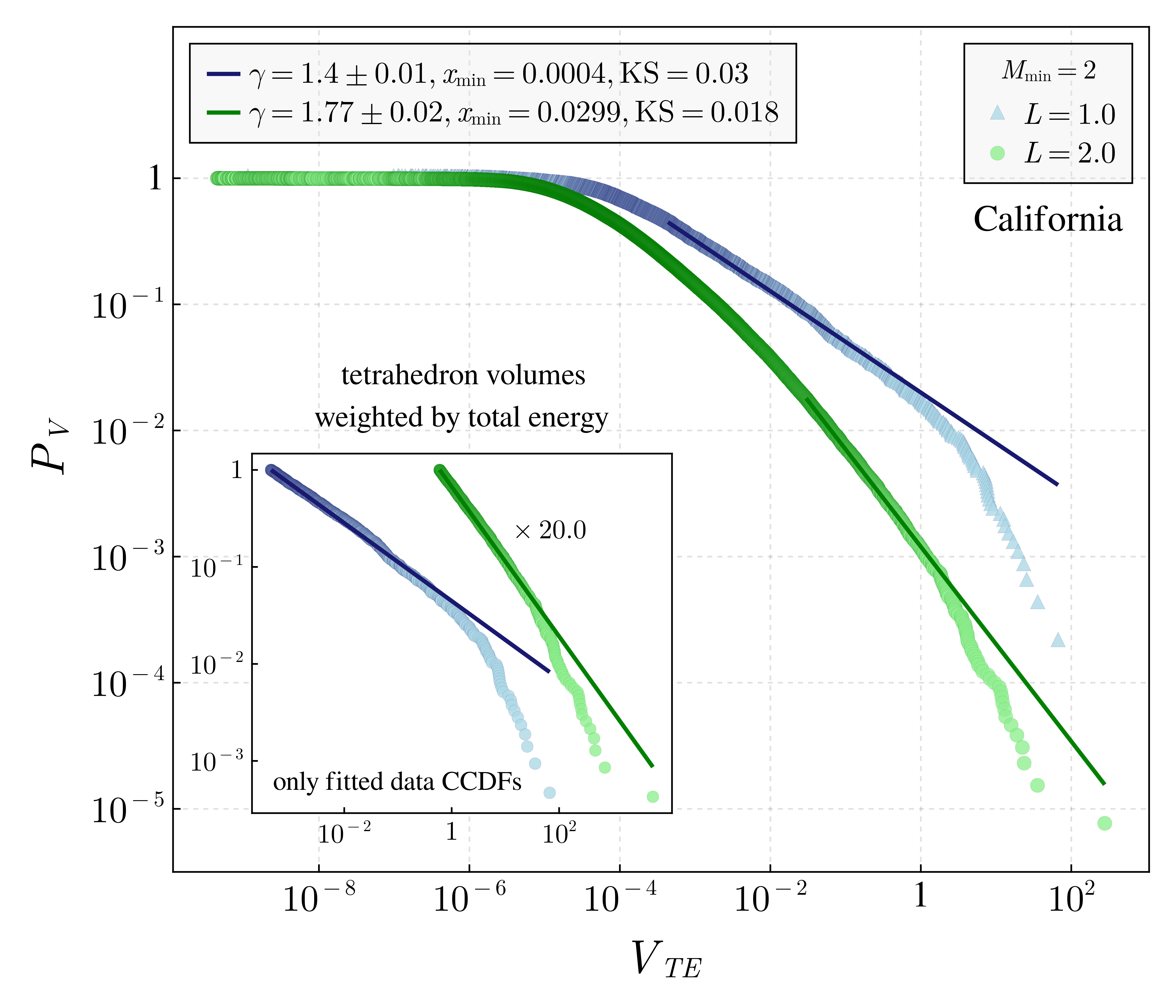}
         \caption{Tetrahedrons}
         \label{fig:cali_tetra}
    \end{subfigure}

    \caption{Complementary cumulative distribution functions for network motifs discovered in networks corresponding to different cube sizes $L$ and a magnitude threshold $M_{\mathrm{min}}=2$ in California. The details are as in Fig. \ref{fig:ro_motifs}. Only the inset in panel (b) presents a shift of the distribution with $L=2.0$ for better visualization.}
    \label{fig:cali_motifs}
\end{figure*}

\begin{figure*}[hbt!]
    \centering
    \begin{subfigure}[b]{0.49\textwidth}
         \centering
         \includegraphics[width=\textwidth]{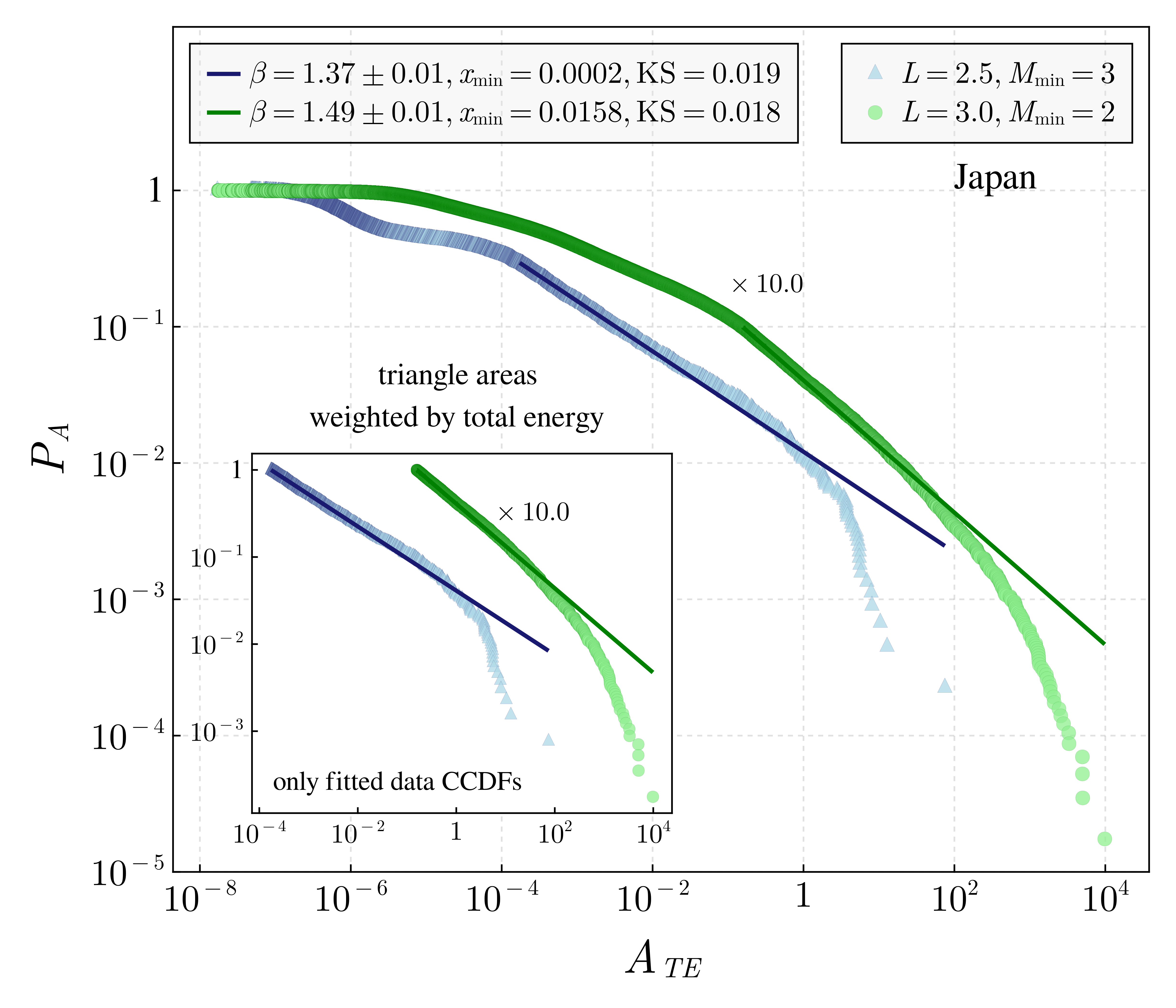}
         \caption{Triangles}
         \label{fig:jap_triangle}
    \end{subfigure}
    \hfill
    \begin{subfigure}[b]{0.49\textwidth}
         \centering
         \includegraphics[width=\textwidth]{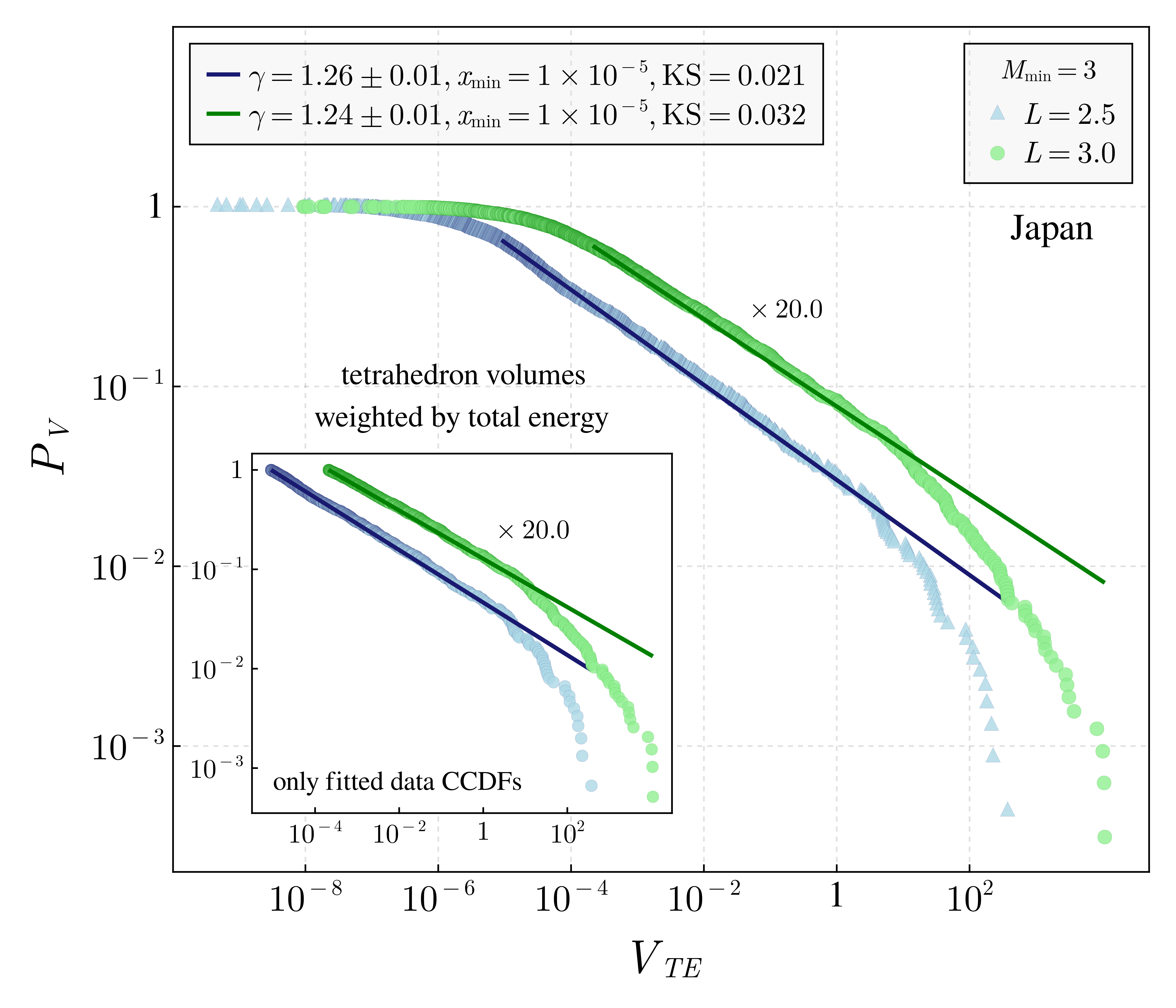}
         \caption{Tetrahedrons}
         \label{fig:jap_tetra}
    \end{subfigure}

    \caption{Complementary cumulative distribution functions for network motifs discovered in networks corresponding to different cube sizes $L$ and magnitude thresholds $M_{\mathrm{min}}$ in Japan. The details are as in Fig. \ref{fig:ro_motifs}. In both panels, we shifted the distribution corresponding to $L=3.0$ by a factor for better visualization.}
    \label{fig:jap_motifs}
\end{figure*}


\subsection{\label{sec:connectivity_analysis}Connectivity analysis}
After creating earthquake networks for different discretizations of the seismic zone, \textit{i.e.}, different values of $L$, we perform a parameter dependency to determine the cell sizes $L$ which correspond to earthquake networks that best approximate the data. We mention here that in this work we do not distinguish between events based on the uncertainty of the epicenter determination and we consider the catalogue as a whole. These results are presented in Fig.~\ref{fig:par_dep} where we notice that for Romania and Italy, $\alpha$ varies greatly for small values of  $L$, then settles at values around $3$ and $2$, respectively, as $L$ increases. In California and Japan, the values of the $\alpha$ exponent show less variation, ranging from $1.5$ to $3.5$. For all regions, the value of $x_{\mathrm{min}}$ is smaller for low $L$, except for California, where we see a substantial decrease in $x_{\mathrm{min}}$ at high $L$, but these fits are poor, as shown by the KS statistic. Fit quality can be easily determined with the help of the accompanying color gradient, good fits generally occurring for values of $L$ around $5 \, \mathrm{km}$ or smaller. Our selection of the values of $L$ for which we report detailed results is based both on the quality of the fit and the value of $x_{\mathrm{min}}$, a small value of the latter being preferred as the resulting power-law component of the distribution function is larger. 

We present detailed connectivity results for three cube lengths $L$ for each seismic zone in Fig. \ref{fig:con_best_fits}. For Romania, in Fig. \ref{fig:ro_con_best_fits}, we notice that robust power-law behavior is seen across only two orders of magnitude, which we suspect is due to the relative small networks, created with a smaller database compared to other regions in this analysis. Varying the cell length $L$ does not produce a significant change to the exponent $\alpha$. In California, Fig. \ref{fig:cali_con_best_fits}, the selected cell sizes for which we obtain good results are smaller compared to the other regions. Robust power-law behavior is observed across three orders of magnitude. Increasing the cell size $L$ induces a decrease in the $\alpha$ exponent. For Italy and Japan, Fig. \ref{fig:it_con_best_fits} and Fig. \ref{fig:jap_con_best_fits}, the power-law behavior is seen over many orders of magnitude. In both regions, an increase in $L$ produces a very small decrease in $\alpha$. The KS error indicator is also small for these regions, indicating a very good fit.

\begin{table}[!t]
\begin{ruledtabular}
\begin{tabular}{ccccc}

                           & \textbf{$L$} & \textbf{$\alpha\pm\sigma$} & \textbf{$\beta\pm\sigma \, (M_{\mathrm{min}})$} & \textbf{$\gamma\pm\sigma  \, (M_{\mathrm{min}})$}  \\ 
\hline
\multirow{4}{*}{Romania}   & $3.0$     & $3.25\pm0.09$     & $1.38\pm0.03 (2)$       & N/A  \\ 
                           & $3.5$     & $3.03\pm0.13$     & $1.41\pm0.03 (2)$       & N/A  \\
                           & $4.5$     & $3.09\pm0.14$     & N/A            & $1.71\pm0.07 (3)$ \\
                           & $5.5$     & $3.01\pm0.15$     & N/A            & $1.92\pm0.04 (3)$ \\
\hline

\multirow{2}{*}{Italy}     & $4.0$     & $1.97\pm0.01$     & $1.54\pm0.01 (2)$  & $1.65\pm0.01 (2)$\\ 
                           & $4.5$     & $1.95\pm0.01$     & $1.52\pm0.01 (2)$  & $1.68\pm0.01 (2)$\\

\hline

\multirow{2}{*}{California}& $1.0$     & $3.04\pm0.06$     & $1.53\pm0.01 (2)$  & $1.40\pm0.01 (2)$ \\ 
                           & $2.0$     & $2.51\pm0.04$     & $1.65\pm0.01 (2)$  & $1.77\pm0.02 (2)$\\
\hline

\multirow{2}{*}{Japan}     & $2.5$     & $2.16\pm0.01$     & $1.37\pm0.01 (3)$  & $1.26\pm0.01 (3)$\\ 
                           & $3.0$     & $2.13\pm0.01$     & $1.49\pm0.01 (2)$  & $1.24\pm0.01 (3)$\\

\end{tabular}
\end{ruledtabular}
\caption{\label{tab:results_info}Results on selected $\alpha$ exponents for connectivity distributions, and $\beta, \gamma$ exponents of the motif analysis.}
\end{table}

\subsection{\label{sec:motifs_analysis}Motifs}
We proceed with the discovery and analysis of the motifs for selected cell sizes $L$, and we also employ a magnitude cut-off for the data $M_{\mathrm{min}}$. Details of the selected parameters are in Table \ref{tab:results_info}. For Romania, the results of our analyses are present in Fig. \ref{fig:ro_motifs}. For triangles \ref{fig:ro_tetra}, an increase in $L$ generates a slight increase in the $\beta$ exponent. For tetrahedrons, \ref{fig:ro_tetra}, we require a bigger magnitude cut-off, $M_{\mathrm{min}}=3$, and observe a similar increase in $\gamma$ when $L$ is larger, similarly to triangles. Please note that the reported results for $\beta$ and $\gamma$ come from different discretizations of the seismic region, a feature which is specific to Romania. This reflects the small size of the seismic database, but it can also reflect some hidden features of the Vrancea seismic zone. In California, Fig. \ref{fig:cali_motifs}, for both triangles and tetrahedrons, an increase in $L$ induces an increase in the $\beta$ and $\gamma$ exponents. The power-law region extends over 5 orders of magnitude in the case of triangles \ref{fig:cali_triangle} and 3 orders of magnitude in the case of tetrahedrons \ref{fig:cali_tetra}.
In Italy, Fig. \ref{fig:it_motifs}, for triangles \ref{fig:it_triangle}, the fits for slightly different $L$ almost coincide, with a $\beta$ exponent of $1.52$. Similarly, in the case of tetrahedrons, \ref{fig:it_tetra}, a small change in $L$ generates a small variation of the $\gamma$ exponent. In Japan, Fig. \ref{fig:jap_motifs}, for triangles \ref{fig:jap_triangle} a small increase in $L$ induces a small increase in $\beta$. The power-law region extends over many order magnitudes. In the case of tetrahedrons \ref{fig:jap_tetra}, we introduce a bigger magnitude cut-off, $M_{\mathrm{min}}=3$, to reduce the time needed to compute the motifs and eliminate from our results the parasitic component caused by small magnitude earthquakes. Please note that the fits almost coincide when changing $L$ from $2.5$ to $3$ km. Our numerical results are summarized in Table \ref{tab:results_info} where we show selected values of $\alpha$ for the connectivity distribution, as well as selected critical exponents for the distribution of motifs across the four seismic regions under scrutiny.

\section{\label{sec:conclusion}Conclusion}

We revisit the databases pertaining to earthquakes in Romania, Italy, and Japan, as well as the California seismic zone in the USA, and show that the earthquake networks used to model these regions have remarkable properties. Largely independent of the discretization of the seismic zones under scrutiny, the earthquake networks used to model them exhibit distributions of node connectivity, three-, and four-event motifs consisting of large power-law components, extending over some orders of magnitude. The analysis of node connectivity is standard, but for three- and four-event motifs, {\it i.e.}, triangles and tetrahedrons, we consider the distributions of the areas and volumes, respectively, weighted in both cases by the total energy released by the earthquakes in the motif. Our analysis relies on maximum likelihood estimation (MLE) and complementary cumulative distribution functions (CCDF) and offers an accurate and rigorous image of the distributions. Our approach is complementary to that obtained from binning methods, which have the particularity that the results depend on the selection of bins. The main message is that the distributions of motifs are similar, all exhibiting a power-law component over some orders of magnitude, and that the scaling exponents of these power-laws can be potentially used to distinguish between different types of seismic zones. This statement can be accurately reinforced only after the current analysis framework is applied to additional seismic areas with similar or different underlying mechanisms.   

Summing up, we have first identified by numerical means the discretization lengths $L$ with which we construct networks that have connectivity distributions well fitted by a power-law distribution. For each seismic zone, we analyzed distributions for three-best cell lengths and found the $\alpha$ exponent to range from $1$ to $\approx 3$. We note that the sizes of the aforementioned three-best cell lengths differ from region to region. In California, for example, our analysis holds best towards smaller cell lengths, $L=1.0, 1.5, 2.0$ $\mathrm{km}$, whereas in the other zones, we find that it holds at around a cell size of $L=5\,\mathrm{km}$. The results for California and Japan are consistent with the findings of \cite{abe04}, which we extend through our distributions of three- and four-event motifs, while for Romania and Italy, our results are new.

The novelty of our analysis relies on the structure of seismic motifs, namely the distribution of three- and four-event motifs, which includes a significant power-law component of scaling exponents $\beta$ and $\gamma$, respectively. We show that these results hold against different discretization lengths and are therefore robust and not an artifact of the underlying discretization. The results presented above include a weighting of the surface of three-event motifs and the volume of the four-event motifs with the total released energy, but the results hold qualitatively also without applying the weights. We also observe that the results are impacted by the size of the available seismic datasets, the most confident modeling of a seismic zone through an underlying network corresponding to the largest available datasets.

Our results reinforce the image of a seismic zone as a self-organized critical system, while at the same time offering a straightforward way of comparing different seismic zones through the aforementioned scaling exponents $\alpha$, $\beta$, and $\gamma$. Moreover, the distribution of three- and four-event motifs can be used as a test bed for new models of seismic activity. 

We expect future studies into earthquake networks to focus on different spatial and temporal scales, potentially connected with the applicability of the fluctuation-dissipation theorem for seismic zones, as well as the study of different complex networks characteristics, such as clustering and community structures \cite{boccaletti2006}, which may reveal interesting correlations within the networks.

\begin{acknowledgments}
We wish to acknowledge fruitful discussions with Virgil B\u{a}ran. This work was supported by the Romanian Ministry of Research, Innovation and Digitalization under Romanian National Core Program LAPLAS VII -- contract no. 30N/2023. The numerical simulations reported here were performed in the computing center of the Faculty of Physics of the University of Bucharest. 
\end{acknowledgments}

\nocite{*}

\bibliography{bibliography}

\end{document}